\documentclass[twocolumn,preprintnumbers,10pt,a4paper]{article}

\usepackage{amsmath}
\usepackage{amssymb}
\usepackage{geometry}
\usepackage[utf8]{inputenc}
\usepackage[T1]{fontenc}
\usepackage{caption}
\usepackage{bbm}
\usepackage{accents}

\usepackage[numbers]{natbib}  

\usepackage{multicol}

\usepackage{mathtools}
\usepackage{graphicx}
\usepackage{dcolumn}
\usepackage{bm}
\newcommand{\be}{\begin{equation}}
\newcommand{\ee}{\end{equation}}
\newcommand{\bd}{\begin{displaymath}}
\newcommand{\ed}{\end{displaymath}}
\newcommand{\BE}{\begin{eqnarray}}
\newcommand{\EE}{\end{eqnarray}}

\newcommand{\bn}{\ensuremath{\mathbf{n}}}

\newcommand{\avg}[1]{\left\langle{#1}\right\rangle}

\usepackage[dvipsnames]{xcolor}
\usepackage[hidelinks]{hyperref}

\newcommand{\change}[1]{#1}

\begin{document}






\date{\today}


\twocolumn[
\begin{center}
{\Large{\textbf{Efficient approximations of transcriptional bursting effects on the dynamics of a gene regulatory network}}}\\\vspace{1cm}

Jochen Kursawe${}^{1}$, Antoine Moneyron${}^{2}$, 
Tobias Galla${}^{3}$ \\~\\
\end{center}

\noindent
${}^{1}$ School of Mathematics and Statistics, University of St Andrews, North Haugh, St Andrews, KY16 9SS, United Kingdom

\noindent
${}^{2}$ Universit\'e Rennes, INRIA, IRMAR UMR 6625, F35000, Rennes, France.

 \noindent
${}^{3}$ Instituto de F\'isica Interdisciplinar y Sistemas Complejos, IFISC (CSIC-UIB), Campus Universitat Illes Balears, E-07122 Palma de Mallorca, Spain
\\
\vspace{0.5cm}
]

\section*{Abstract}
Mathematical models of gene regulatory networks are widely used to study cell fate changes and transcriptional regulation. When designing such models, it is important to accurately account for sources of stochasticity. However, doing so can be computationally expensive and analytically untractable, posing limits on the extent of our explorations and on parameter inference. Here, we explore this challenge using the example of a simple auto-negative feedback motif, in which we incorporate stochastic variation due to transcriptional bursting and noise from finite copy numbers. We find that transcriptional bursting may change the qualitative dynamics of the system by inducing oscillations when they would not otherwise be present, or by magnifying existing oscillations. We describe multiple levels of approximation for the model in the form of differential equations, piecewise deterministic processes, and stochastic differential equations. Importantly, we derive how the classical chemical Langevin equation can be extended to include a noise term representing transcriptional bursting. This approximation drastically decreases computation times and allows us to analytically calculate properties of the dynamics, such as their power spectrum. We explore when these approximations break down and provide recommendations for their use. Our analysis illustrates the importance of accounting for transcriptional bursting when simulating gene regulatory network dynamics and provides recommendations to do so with computationally efficient methods.

\section*{Media summary}
Dynamic changes in gene expression govern many crucial processes in living cells, including fate transitions and growth. Mathematical models of gene expression help us understand these dynamics. Such models can have a large computational cost, requiring efficient approximations. However, typical approximations ignore a major source of random variation: the transcription of genes in bursts. We provide approximations of gene regulation that account for bursting and demonstrate their accuracy on multiple examples. Our examples highlight that bursting leads to drastic differences in the behaviour of gene regulatory networks, inducing large oscillations in expression and fate transitions that would not otherwise occur.

\section*{Keywords} Gene regulatory networks, gene expression oscillations, transcriptional bursting, fast-switching environments, piecewise deterministic pseudo-Markov processes, chemical Langevin equations

\section{Introduction}
Gene regulatory networks are at the heart of many developmental and physiological processes, and are a key component of cell fate decisions \cite{davidson}. 
The expression of proteins participating in gene regulatory networks, i.e. transcription factors, is inherently dynamic. A fundamental example of dynamic transcription factor expression are gene expression oscillators, such as the circadian clock \cite{patke,russel,papagiannakis,kalsbeek}, or rapid ultradian oscillators governing the formation of somites \cite{miao}.
Rapid ultradian gene oscillators also encode information necessary for cell-state transitions, such as the differentiation of neural progenitor cells into neurons or glial cells \cite{kageyama,philips}. 
 
It is surprising that dynamic signatures of gene expression, such as oscillations, are correctly established and interpreted when considering the prevalence of stochastic effects \cite{raj}. The copy numbers of protein and mRNA molecules in gene regulatory networks are often sufficiently small that simple mathematical approaches based on deterministic rate equations fail to provide even a qualitatively accurate description \cite{kaern, schlicht, barrio, bratsun}. For small copy numbers the intrinsic stochasticity of transcription and translation processes and of the degradation of protein and mRNA cannot be neglected. This raises the key question whether noise is detrimental to the functioning of gene regulatory processes, or if indeed, stochasticity can be exploited \cite{eldar}. An example of the latter is \emph{stochastic amplification}, in which a stochastic dynamical system may show persistent oscillations in circumstances when its corresponding deterministic system does not \cite{alan1,alan2,galla2009}. 

Gene regulatory networks are subject to other types of stochasticity, beyond copy-number noise. For example, conditions external to the cellular environment may vary in time or cell divisions might be uneven \cite{baptista, huh}. 
One additional source of randomness is transcriptional bursting, describing the phenomenon that mRNA production is discontinous in time and occurs in bursts of transcriptional activity. The size and timing of such bursts varies stochastically \cite{porello,larsson}. 

Mathematical models are a widely used tool to study and and identify gene regulatory networks \cite{rand,chai,pratapa}, and their description through the law of mass action \cite{ferner,alon,tyson} can easily be extended to account for copy-number noise through the Gillespie stochastic simulation algorithm \cite{gillespiessa} or the chemical Langevin equation \cite{gillespieCLE,vankampen}. It is an ongoing challenge to understand how gene regulatory networks integrate other types of noise, such as transcriptional bursting, to robustly establish dynamical signatures like oscillations and cell fate transitions.
%


To mathematically describe experimental observations on transcriptional bursting of a single gene (i.e. not considering responses within a wider gene regulatory networks), the so-called telegraph model has been widely used. In this simplest model of transcriptional bursting, the gene randomly switches between states of active and inactive transcription \cite{Ko,munsky}. Considerable effort has been expended to parameterise versions of the telegraph model using experimental data and a recent overview of such approaches is for example available from Luo et al. \cite{Luo2023}. Typically, the comparison between models and data is performed using steady-state distributions of mRNA copy numbers, which can be predicted from the model and experimentally measured using scRNA-seq data \cite{Luo2022} or smFISH \cite{sepulveda}. In some cases, bursting dynamics are directly observed through live-imaging and used to identify parameters of telegraph models and related motifs \cite{fu,so,suter,li}.

In addition to these efforts at the level of a single gene, many studies investigate the effects of transcriptional bursting into wider gene regulatory networks. An example are inference approaches that take regulatory interactions into account when inferring properties of transcriptional bursting \cite{schuh,herbach}. Recent work by \cite{gupta}, points out that co-fluctuations between transcription factors and their targets may inform inference, and earlier work by \cite{zhu} and \cite{rajala} have discussed in detail how molecular processes of transcriptional as well as translation bursting may be incorporated into stochastic descriptions of gene regulatory networks. These efforts often rely on exact simulations of the stochastic dynamics, which can be computationally costly especially when copy numbers of the involved molecules are large. In addition, exact description of the stochastic processes involved are typically not analytically tractable. These shortcomings limit the potential of stochastic simulations in studying gene regulatory networks. 

Here, we investigate the delayed auto-negative feedback motif, in which we incorporate the effects of transcriptional bursting as well as noise due to finite copy numbers. Using the example of this motif, we introduce multiple levels of approximations of gene regulatory dynamics valid in the individual or joint limits of large copy numbers or fast bursting. As part of this effort, we extend the classical Chemical Langevin equation to account for Gaussian effects of transcriptional bursting in gene regulatory networks. We show that stochastic variation due to transcriptional bursting can enable oscillations when they are not otherwise present, or enhance oscillations if they are already present. Our approximations allow us to analytically determine the power spectrum of the oscillations, and investigate for what system sizes and burst frequencies our approximation become invalid. Our derivations of the extended chemical Langevin equation are provided in general form. These equations are approximations that drastically reduce computation times, and are applicable to any gene regulatory network and many models of transcriptional bursting.


The remainder of the paper is organised as follows. In the model description of section~\ref{sec:model} we define the model of the auto-repressive network of a single gene, explicitly including finite copy numbers and transcriptional bursting, as well as transcriptional delay. We demonstrate that our model simplifies to previously published versions of the model in the limit of infinitely fast transcriptional bursting, and discuss the alternative limiting case where copy-number noise is absent. In the results section ~\ref{sec:results} we first discuss the effect of transcriptional bursting on the system dynamics before presenting our main analytical results. Through a systematic expansion in the inverse noise strength we are able to calculate the Fourier spectra of oscillations induced by copy number noise and/or noise due to bursting. We proceed to test these theoretical predictions against simulations and assess the validity of the various reduced models for different regimes of transcriptional bursting and intensity of copy-number noise. We finally include a discussion in section~\ref{sec:summary}. Further details of the mathematical analysis are collated in the supplementary material.

\begin{figure}
    \centering
   \includegraphics{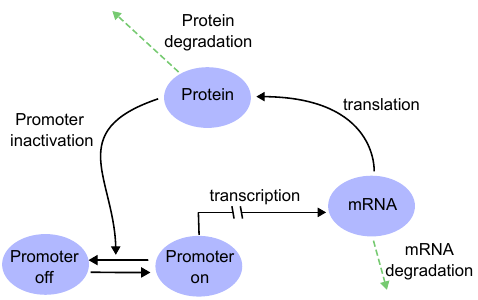}
    \caption{\textbf{Schematic of the transcriptional auto-repressive feedback loop.} Our model includes the effect of a promoter which can be in an active or an inactive state (on/off), as well as the dynamics of mRNA and protein. These are subject to transcription, translation, degradation. The promoter can be inactivated by the presence of protein. The two dashes on the arrow for transcription denote the presence of a transcriptional delay $\tau$.}
    \label{fig:circuit}
\end{figure}

\section{Model description}\label{sec:model}
\subsection{Gene regulatory network with negative auto-regulation and transcriptional bursting}
We start by introducing a simple model of a transcription factor subject to auto-negative feedback (figure \ref{fig:circuit}). Importantly, this model accounts for transcriptional bursting as well as stochastic variation due to low copy numbers. The model extends previous work \cite{monk,galla2009} and includes the effects of transcription, translation, and degradation of mRNA and protein. Motivated by the classical telegraph model of transcription, the promoter of the gene can assume two states, one state in which transcription is active with a constant rate, and one state in which the transcription does not occur. The promoter switches between these two states stochastically, and the rate of switching into the off-state depends on the amount of protein present in the system. This simulates the auto-negative feedback, whereby a high abundance of protein can inhibit transcription.

Mathematically, our model comprises a set of chemical reactions between protein and mRNA molecules, and their interactions with the promoter state. 
The number of each type of particle is discrete. We write $n_P$ for the number of protein molecules, and $n_M$ for the number of mRNA molecules
in a cell. They relate to their concentrations $M$ and $P$ via $n_M = M V$ and $n_P=P V$, where $V$ 
 is the nuclear volume. For convenience, we write $V=\Omega V_0$, where we call $\Omega$  the \textit{system size} in-line with \change{chapter 10 of \cite{vankampen}}. By varying the size of the nucleus, $\Omega$ controls how many molecules may be expected in the system at constant concentrations.
Throughout the paper, we will use parameter values identified in \cite{manning_et_al2019}, and we choose $V_0=523fL$, which is the average nuclear volume in that publication. We will measure concentrations in concentration units `cu', which correspond to one molecule per volume $V_0$.

Following \cite{barrio,bratsun} our model is 
\BE
G_{\text{ON}} &\stackrel{\lambda (P/P_0)^h}{\longrightarrow}& G_{\text{OFF}}, \nonumber \\
G_{\text{OFF}} &\stackrel{\lambda }{\longrightarrow}& G_{\text{ON}}, \nonumber \\
M&\stackrel{\mu_M}{\longrightarrow}& \varnothing, \nonumber \\
P&\stackrel{\mu_P}{\longrightarrow}& \varnothing, \nonumber \\
\varnothing &\stackrel{\alpha_M \sigma}{\Longrightarrow}& M, \nonumber \\  
M&\stackrel{\alpha_P}{\longrightarrow}& M+P, \label{eq:full_reactions}
\EE
where $G_{\rm ON}$ and $G_{\rm OFF}$ stand for the promoter states in which transcription of the gene are ON and OFF respectively. We refer to the parameter $\lambda$ in the first two lines in (\ref{eq:full_reactions}) as the \textit{bursting parameter.} It controls the time scale of switching between the two promoter states. Note, that the reaction rate from the promoter OFF into the ON state is in the literature often referred to as burst frequency, which is equal to $\lambda$ in our parameterisation \cite{larsson,suter}. 
$h$ is a so-called Hill coefficient. We refer to the fixed model parameter $P_0$ as the \textit{repression threshold}, which controls the amount of protein required to inhibit transcription. 
The third and fourth reactions describe degradation of mRNA and protein, respectively, with degradation rates $\mu_M$ and $\mu_P$.
For convenience, we introduce the mathematical variable $\sigma(t)$ to indicate if the promoter is in the ON state [$\sigma(t)=1$] or the OFF state [$\sigma(t)=0$] at any one time $t$. 
We draw attention to the rate $\alpha_M \sigma$ for the penultimate reaction, which describes transcription. This reaction rate indicates that a transcription process can only be initiated at a given time $t$ if the promoter is on the ON state [$\sigma(t)=1$]. The rate of transcription is given by the basal transcription rate $\alpha_M$. Note, that an alternative way to write the transcription reaction without the use of the $\sigma$ variable would be
\begin{equation}
G_\mathrm{ON}  \stackrel{\alpha_M}{\Longrightarrow} G_\mathrm{ON} + M.
\end{equation}
Here, we instead use the notation in equation \eqref{eq:full_reactions}, as the variable $\sigma$ will be useful in our calculations.

The resulting mRNA molecule of a transcription event that is triggered at time $t$ enters the system at time $t+\tau$. This delay is indicated by the double arrow in the reaction system and simulates the time required for transcription and subsequent transport of the mRNA out of the nucleus \cite{monk,lewis,alberts}.
The final reaction describes the translation of mRNA into protein, with rate $\alpha_P$. 

The number of reactions of each type that occur in each time interval are either proportional to $\Omega$, $n_P$ or $n_M$. As a consequence, the copy numbers of mRNA and protein particles in the system are proportional to the system size parameter $\Omega$. We can use this parameter to control the strength of copy number noise in our analysis, as stochasticity in the rate of each reaction will decrease with the system size.

We highlight that our calculations are consistent with other mechanisms that upregulate the expected number of molecules in the system by a factor $\Omega$. In fact, biological cells can have vastly different expression levels for individual proteins without an increased size of the nucleus. Scaling by $\Omega$ as in our system can be used to model such increases, although we point out that $P$ and $M$ in this case are not technically concentrations. Rather, increases in $\Omega$ would correspond to an effective change of the unit of concentration towards higher density.  In our model, that would be achieved by increasing the basal transcription rate and the repression threshold by a factor $\Omega$, as this would be mathematically equivalent to an increase in volume as before.

We chose the form of the transition rate into the promoter off-state of $\lambda(P/P_0)^h$. This reaction rate reflects the assumption that, biophysically, a promoter requires binding with $h$ protein molecules to switch into its OFF state. This can be seen as follows.

Considering a promoter which can bind $h$ protein molecules, the promoter may assume an unbound state $G_\mathrm{OFF}$ and a bound state $G_\mathrm{ON}$, where the latter corresponds to the state in which all $h$ binding sites are occupied. The reaction kinetics of promoter binding and unbinding are then as follows,
\be\label{eq:bu}
G_\mathrm{ON} + h P \xrightleftharpoons[k_2]{k_1} G_\mathrm{OFF},
\ee
where $k_1$ and $k_2$ are binding constants. The left-hand side describes the unbound state and $h$ free protein molecules. On the right-hand side, all $h$ protein molecules are bound. In this reaction system, the rate of the reaction into the OFF state is given by 
\begin{equation}
k_\mathrm{OFF} = k_1 P^h
\end{equation}
and the reaction into the $ON$ state is given by
\begin{equation}
k_\mathrm{ON} = k_2.
\end{equation}
We now define $k_1 = \lambda/P_0^h$ and $k_2 = \lambda$. Under these definitions, reaction \eqref{eq:bu} is equivalent to the first two reactions in equation \eqref{eq:full_reactions}. \change{In doing so, we parameterise the promoter dynamics such that both transition rates between the OFF and ON states depend on $\lambda$. As a consequence, the mean probability to be in the ON or OFF state does not depend on $\lambda$, keeping the time-averaged transcription rate constant as we vary the bursting parameter $\lambda$. This allows us to isolate effects of promoter dynamics from those of changes in transcription rates.}

\subsection{Our model simplifies in the limit of infinitely fast transcriptional bursting}
We next consider the limit of infinitely fast bursting parameter, $\lambda\rightarrow\infty$.
We write $p_\mathrm{ON}$ and $p_\mathrm{OFF}$ for the probabilities to find the promoter in its unbound and bound states, respectively. Reflecting the limit $\lambda\rightarrow \infty$, we assume that switching between the promoter states is sufficiently fast to reach an equilibrium while a given concentration $P$ of protein molecules is present in the system. In this equilibrium of promoter binding and unbinding we have
\be
\lambda~ p_\mathrm{ON}~ \left(\frac{P}{P_0}\right)^h = \lambda~ p_{OFF}.
\ee
in addition to 
\begin{equation}
p_\mathrm{ON} + p_\mathrm{OFF} = 1.
\end{equation}
Hence, in this stationary state, the probability to find the promoter in its bound OFF state is 
\begin{equation}
p_\text{OFF}=\frac{ (P/P_0)^h}{1+ (P/P_0)^h}.
\end{equation} 
The probability to find the promoter in the unbound ON state is 
\be\label{eq:p_unbound}
p_{\text{ON}} =\frac{1}{1+ (P/P_0)^h}=: f(P).
\ee

The average transcription rate is then given by 
\begin{equation}
\alpha_M p_\mathrm{ON} + 0~p_\mathrm{OFF} = \alpha_M f(P) = \alpha_M\frac{1}{1+ (P/P_0)^h}.\label{eq:hill}
\end{equation}
If the promoter switching is fast relative to the remaining reactions, i.e. when $\lambda \rightarrow \infty$, fluctuations around this average value will diminish, and the system can be approximated as
\BE
M&\stackrel{\mu_M}{\longrightarrow}& \varnothing \nonumber \\
P&\stackrel{\mu_P}{\longrightarrow}& \varnothing\nonumber \\
\varnothing &\stackrel{\alpha_M f}{\Longrightarrow}& M \nonumber \\  
M&\stackrel{\alpha_P}{\longrightarrow}& M+P. \label{eq:reactions}
\EE
 The function $f$ is a decreasing function of the protein concentration $P$, reflecting auto-repression and the degree of non-linearity of the function, i.e. its sensitivity to changes in protein molecule numbers is regulated by the Hill coefficient $h$. As before, the double arrow in equation \eqref{eq:reactions} denotes the presence of the transcriptional delay $\tau$. These reactions were previously introduced by \cite{monk} and they have since then been used widely as a template model for auto-repressive feedback \cite{barrio,philips,galla2009,goodfellow}. Our model in figure \ref{fig:circuit} and equation \eqref{eq:full_reactions} is intentionally designed so that the promoter behaves as in the widely used telegraph model for transcriptional bursting, while also achieving the typically used Hill function \eqref{eq:hill} for the transcription rate for the case of infinitely fast bursting. This allows us to analyse how the dynamics of the model \eqref{eq:reactions} are modified if transcriptional bursting is taken into account.  
The description in terms of the Hill function is valid, provided the promoter switches between the ON and OFF states much more quickly than the remaining time scales in the reaction system. 

A deterministic description of these dynamics is given by the following rate equations (see also \cite{monk,galla2009}) for the concentrations $M$ and $P$,
\BE
\frac{d}{dt} M(t)&=&\alpha_M~f[P(t-\tau)]~-\mu_M~M(t), \nonumber \\
\frac{d}{dt} P(t)&=&\alpha_P~M(t)-\mu_P~P(t) \label{eq:monk},
\EE

We note the delay term in the first equation. This description does not describe fluctuations due to finite copy numbers or transcriptional bursting, and is formally only valid in the joint limit of $\Omega\to\infty$ and $\lambda\to\infty$.

\begin{table*}[t]
    \centering
    \begin{tabular}{c|c|c}
     &{\bf finite copy numbers}  & {\bf infinite copy numbers } \\[0.5em] 
      & ($\Omega<\infty$)  & ($\Omega\to\infty$) \\[0.5em] 
      \hline
    {\bf finite transcriptional }  & copy number noise & only transcriptional bursting noise \\[0.5em] 
    {\bf bursting parameter} ($\lambda<\infty$)& and transcriptional bursting noise & piecewise deterministic dynamics\\[0.5em] 
    & full model, equation~(\ref{eq:full_reactions}) & equation~(\ref{eq:pdmp})\\[0.5em]
    \hline
     {\bf infinite transcriptional }  & only copy-number noise & no noise  \\[0.5em]
     {\bf bursting parameter} ($\lambda\to\infty$) & reaction system in equation~(\ref{eq:reactions}) & delay differential equations (\ref{eq:monk}) \\[0.5em]
    \end{tabular}
    \caption{\textbf{Overview of the different descriptions of the combined system of transcriptional bursting and protein-mRNA dynamics} Further details are available in the main text.}\label{tab:models}
\end{table*}

\subsection{Limiting cases distinguish sources of noise in the system: copy-number noise vs. noise due to transcriptional bursting}
We can distinguish between two types of randomness in the model in equation~(\ref{eq:full_reactions}). The first is copy number noise. By this we mean the intrinsic stochasticity in the protein and mRNA dynamics, described by the last four reactions in (\ref{eq:full_reactions}). This randomness is present even at fixed promoter state $\sigma$, and becomes more pronounced when there are only small numbers of molecules in the system. More precisely, the amplitude of intrinsic noise scales as $\Omega^{-1/2}$ for large values of the system-size parameter $\Omega$ \change{(\cite{vankampen}, chapter 10, or \cite{alan1}, and \cite{alan2})}. This is a consequence of the central limit theorem. Broadly, the mean number of reactions in the system per unit time is proportional to $\Omega$, and the standard deviation of the number of reactions is of order $\Omega^{1/2}$. Relative fluctuations therefore scale as $\Omega^{-1/2}$. 

The second type of noise arises from transcriptional bursting, induced by the random switching between the ON and OFF states in the first two reactions in (\ref{eq:full_reactions}). The promoter dynamics operate on time scales which are set by the bursting parameter $\lambda$. 
For increasingly pronounced time-scale separation, i.e., for very fast transcriptional burst frequencies ($\lambda\gg 1$), the transcriptional bursting effects average out, and thus the effects of transcriptional burstiness diminish. Using again the central limit theorem the relative fluctuations due to transcriptional bursting scale as $\lambda^{-1/2}$. The reactions in equation~(\ref{eq:full_reactions}) describe the full model, for general values of $\Omega$ and $\lambda$. This setup therefore captures both copy number noise and noise due to transcriptional bursting.

Having considered the limit $\lambda\rightarrow\infty$, equation \eqref{eq:reactions}, in which only copy-number noise and no bursting noise is present, and the joint limit $\lambda\rightarrow\infty,~\Omega\rightarrow\infty$, equation \eqref{eq:monk} in which neither noise is present, it remains to cover the limit $\Omega\rightarrow\infty$, in which bursting noise is present, but copy-number noise is not. 
In this limit, the dynamics of protein and mRNA may be described by ordinary differential equations. However, for general values of $\lambda$, the promoter dynamics remains stochastic. This leads to a description similar to what is known as a piecewise deterministic Markov process \cite{pdmp_davis,hufton_pre_1}. Specifically, in this limit one has
\BE
\frac{d}{dt} M(t)&=&\alpha_M~\sigma(t-\tau)-\mu_M~M(t),  \nonumber \\
\frac{d}{dt}P(t)&=&\alpha_P~M(t)-\mu_P~P(t)\label{eq:pdmp}
\EE
for the dynamics of mRNA and protein. The switching of the promoter state is as before, that is transitions from $\sigma=0$ to $\sigma=1$ occur with the bursting parameter $\lambda$, and the promoter switches off ($\sigma=1 \to \sigma=0$) at time $t$ with rate $\lambda [P(t)/P_0]^h$. We refer to equation \eqref{eq:pdmp} as the equations for the piece-wise deterministic pseudo-Markov process (PDPMP). This approximation has recently been analysed in the stationary regime \cite{wang2023}.

We note the term $\alpha_M \sigma(t-\tau)$ in the first equation in (\ref{eq:pdmp}). This describes the process mRNA production, subject to transcription delay. Production of mRNA molecules at time $t$ only occurs if the promoter was on the ON state ($\sigma=1$) at time $t-\tau$. If on the other hand $\sigma(t-\tau)=0$ then no new mRNA molecules enter the system at time $t$.


The different descriptions of the system are summarised in Table~\ref{tab:models}.

\section{Results}\label{sec:results}
\subsection{Oscillatory gene expression can be induced and amplified by transcriptional bursting}
\begin{figure*}
    \centering
    \includegraphics{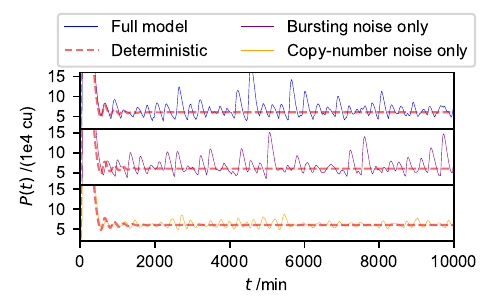}
    \caption{\textbf{Oscillatory gene expression can be induced and amplified by transcriptional bursting}. 'Full model' refers to  equation~\eqref{eq:full_reactions}, `bursting noise only' refers to \eqref{eq:pdmp}, valid in the limit $\Omega\to\infty$ and $\lambda$ remaining finite and `copy-number noise only' refers to equation \eqref{eq:reactions} for infinitely fast transcriptional burst frequencies ($\lambda\to\infty$, $\Omega$ finite). The deterministic model refers to equation \eqref{eq:monk}, valid in the combined limit $\Omega\to\infty$ and $\lambda\to\infty$. Parameters are according to figure~6e of \cite{manning_et_al2019}. In detail they are: $\alpha_M=39.93{\rm cu/min}$, $\alpha_P=21.56{\rm  min}^{-1}$, $\mu_m=(\ln\,2/30) {\rm min}^{-1}$, $\mu_p=(\ln\,2/90) {\rm min}^{-1}$, $h=4.78$, $P_0=24201.01$cu (concentration unit), and $\tau = 33 {\rm min}$. We use $\Omega = 1$ for the full model and the copy-number noise only model, whereas the remaining models assume $\Omega=\infty$. Similarly, we use $\lambda = 0.1$min$^{-1}$ for the full model and the bursting-noise only model, where the remaining models assume $\lambda=\infty$. Details of the computational implementation are available in supplementary section S3.}
\label{fig:timeseries_noise}
\end{figure*}

Our model \eqref{eq:full_reactions} extends the widely used model of auto-negative feedback \eqref{eq:reactions} to take transcrptional bursting into account. This allows us to ask how transcriptional bursting may affect the dynamics of the feedback loop. To this end, we simulate the full model \eqref{eq:full_reactions}, including bursting noise and copy-number noise, as well as the limiting cases with only one type of noise (equations \eqref{eq:reactions} and \eqref{eq:pdmp}), and no noise (equation \eqref{eq:monk}). We choose experimentally informed parameter values for $\mu_P$ and $\mu_M$, as well as the parameters $\alpha_M, \alpha_P, P_0$ and the transcription delay $\tau$ from mouse neural progenitor cells reported in \cite{manning_et_al2019} and detail our simulation algorithms in supplementary section S3. As an initial example, we choose $\Omega = 1$ and bursting parameter $\lambda=0.1$min$^{-1}$. Intriguingly, we find that the presence of transcriptional bursting can strongly enhance oscillations in the system (figure \ref{fig:timeseries_noise}).

Specifically, we plot in figure \ref{fig:timeseries_noise} the concentration of protein molecules $P$ as a function of time for each model. The dashed line in each panel of figure \ref{fig:timeseries_noise} is obtained in the absence of either type of noise, i.e., from the deterministic model in equation~(\ref{eq:monk}). As seen in the figure, the concentration of protein shows decaying oscillations and settles to a stable fixed point. Thus, without noise there is no persistent oscillatory behaviour for this parameter set.

Considering simulations of the full model, comprising both copy number noise and transcriptional bursting (equation~\eqref{eq:full_reactions}), we find that the time series shows persistent oscillations. Given that these oscillations are absent in the fully deterministic model (dashed line), the oscillatory behaviour must be noise-induced.  

Next, we consider the piecewise deterministic model obtained in the limit of infinite copy numbers (equation~\eqref{eq:pdmp}), i.e. the model in which transcriptional bursting occurs, but copy-number noise is neglected (`bursting noise only' in figure \ref{fig:timeseries_noise}). We again observe noise-induced oscillatory behaviour with a slightly lower amplitude than in the full model. 

Finally, we consider the limit of infinitely fast bursting dynamics and finite copy numbers, as in equation \eqref{eq:reactions}, (`only copy-number noise' in figure \ref{fig:timeseries_noise}). We again observe noise-induced oscillations. As these oscillations represent the result of an amplification of copy-number noise only, they are similar to what was reported in other cases of stochastic amplification \cite{guemez,alan1,alan2,galla2009,manning_et_al2019}. Notably, the amplitude of the oscillations in this case is considerably lower than that in the full model or the `bursting noise only' model, illustrating that the presence of transcriptional bursting changes the qualitative dynamics of the system by amplifying these oscillations.


\subsection{Variation due to transcriptional bursting can be approximated with stochastic differential equations}\label{sec:theory}
Our results in figure 2 indicate the importance of considering transcriptional bursting when investigating the dynamics of gene regulatory networks. Our analysis relied on simulations of the full reaction system in equation \eqref{eq:full_reactions}. These reactions allow us to simulate the system dynamics under the presence of transcriptional bursting and copy-number noise. They can be simulated using the Gillespie stochastic simulation algorithm \cite{gillespiessa,cai}, and mathematically described using the master equation formalism \change{(see e.g. \cite{vankampen}, chapter 7)}, or, on our case, variations of it that account for transcriptional delay \cite{galla2009}. However, master equations are notoriously hard to solve, and for analytical derivations chemical Langevin equations are widely used \cite{gillespieCLE,galla2009}. Chemical Langevin equations have the added benefit of requiring orders of magnitude fewer calculations in each simulation than the Gillespie simulation algorithm. 

Unfortunately, the chemical Langevin equation cannot readily be applied to systems with switching promoter states. Here, we build on theory developed by \cite{hufton_pre_1} to demonstrate how the chemical Langevin equation formalism can be applied to incorporate Gaussian approximations of transcriptional bursting. This latter approximation will be valid in the limit of large but finite values of $\lambda$.


To derive an effective stochastic differential equation encompassing both copy number and bursting noise we adapt the procedure first used by Gillespie in \cite{gillespieCLE}. More precisely, we discretise time, and then require that the time step is sufficiently small such that reaction rates and burst frequencies remain constant during each step. We simultaneously assume that the time step is large enough such that the number of reactions in each time step follows approximately a Gaussian distribution. These two assumptions can only be simultaneously fulfilled in system with large system sizes $\Omega$ and large burst frequencies $\lambda$. Under these assumptions it is possible to analytically calculate the mean and variance of the Gaussian increments at each timestep, and the definition of a stochastic differential equation emerges. Our derivation is described in full detail in Section~S1 of the  supplementary material, where we provide formulas relating to model \eqref{eq:full_reactions} as well as for more general gene regulatory networks. 

We obtain the following extended chemical Langevin equations,
\BE\label{eq:CLE_main}
\dot M &=& f[P(t-\tau)]-\mu_M M +\xi_M(t), \nonumber \\
\dot P &=& \alpha_P M - \mu_P P +\xi_P(t),
\EE
with Gaussian white noise variables $\xi_M(t)$ and $\xi_P(t)$, each of mean zero, and with 
\BE
\avg{\xi_M(t)\xi_M(t')}&=&\frac{\alpha_M^2}{\lambda}\frac{2[P(t-\tau)/P_0]^h}{\big(1+[P(t-\tau)/P_0]^h\big)^3} \delta(t-t') \nonumber\\
&&\left.\hspace{-5em} + \frac{1}{\Omega} \big[f[P(t-\tau)] + \mu_M M\big]\right.\delta(t-t') \nonumber \\
\avg{\xi_P(t)\xi_P(t')}&=&\frac{1}{\Omega} \left[\alpha_P~M(t)+\mu_P~P(t)\right]\delta(t-t') \nonumber \\
\avg{\xi_M(t)\xi_P(t')}&=&0. \label{eq:noise_cov_main}
\EE
We note two contributions to $\avg{\xi_M(t)\xi_M(t')}$, one proportional to $1/\Omega$, and representing copy-number noise, and a second proportional to $1/\lambda$, representing noise due to transcriptional bursting. Protein production and decay are not dependent on the promoter state, and hence the noise variable $\xi_P$ originates purely from copy-number noise. Its variance is therefore proportional to $1/\Omega$. The last expression in equation~(\ref{eq:noise_cov_main}) indicates that there are no correlations between $\xi_M$ and $\xi_P$. This is because no reaction in equation~(\ref{eq:full_reactions}) changes both the number of protein molecules and that of mRNA molecules.

The extended chemical Langevin equation \eqref{eq:CLE_main} provides a possible approximation of the full model in equation \eqref{eq:full_reactions}. The dynamics of the chemical Langevin equation are dictated by two coupled stochastic differential equations. This is a drastic simplification of the full reaction system, which is described by an extended chemical master equation that accounts for the presence of a delay \cite{galla2009,brett}. Mathematically, this chemical master equation constitutes an infinitely large set of coupled ordinary differential equations \change{(\cite{vankampen}, chapters 5 and 7)}. Nonetheless, the chemical Langevin equation is non-linear, making it analytically intractable. An analytically tractable equation can be obtained by linearising. This involves one further step of approximation and results in the so-called linear-noise approximation.

Our formulation of the linear-noise approximation applies to parameters for which the deterministic system in equation~\eqref{eq:monk} tends to a stable fixed point $(M^*, P^*)$. 
We linearise about this fixed point by writing
\BE
M(t)&=& M^*+m(t), \nonumber \\
P(t)&=& P^*+p(t).
\EE
We note that $m(t)$ and $p(t)$ describe fluctuations of particle concentrations (as opposed to fluctuations of particle numbers) about the deterministic fixed point. 

Assuming that $m(t)$ and $p(t)$ are small quantities, we then obtain the following linear stochastic delay differential equations,
\BE
\dot m &=& \alpha_M~f'(P^*)~p(t-\tau)-\mu_M~m(t) +\zeta_M(t) \nonumber \\
\dot p &=& \alpha_P~m(t) - \mu_P~p(t) +\zeta_P(t), \label{eq:lna_main}
\EE
where 
\be\label{eq:F_prime}
f'(P)=\frac{df}{dP}=\frac{-h}{P_0}\frac{1}{[1+(P/P_0)^h]^2}\left(\frac{P}{P_0}\right)^{h-1}.
\ee
A key element of the linear-noise approximation is to replace the second moments of the noise variables in the chemical Langevin equation, equations~(\ref{eq:noise_cov_main}), with the values of these expressions at the deterministic fixed point. This leads to Gaussian white noise variables $\zeta_M$ and $\zeta_P$ with properties
\BE
\avg{\zeta_M(t)\zeta_M(t')}&=&\sigma_M^2~\delta(t-t'),\nonumber \\
\avg{\zeta_P(t)\zeta_P(t')}&=&\sigma_P^2~\delta(t-t'),\nonumber \\
\avg{\zeta_M(t)\zeta_P(t')}&=&0, \label{eq:zetazeta}
\EE
where
\BE
\sigma_M^2&=&\frac{1}{\Omega}\left[\alpha_M ~f(P^*)+\mu_M M^*\right] \nonumber \\
&&+ \frac{1}{\lambda}\frac{2[P^*/P_0]^h}{\left(1+[P^*/P_0]^h\right)^3}\alpha_M^2,\nonumber \\
\sigma_P^2&=&\frac{1}{\Omega}\left[\alpha_P M^*+\mu_P P^*\right].\label{eq:sigma}
\EE

\begin{figure*}[t]
    \centering
\includegraphics{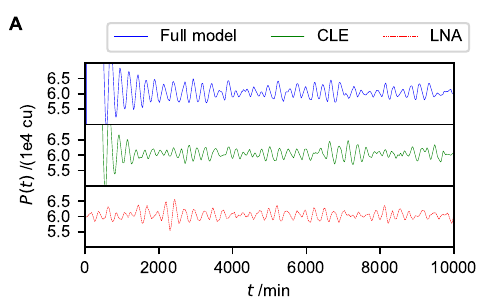}      
\includegraphics{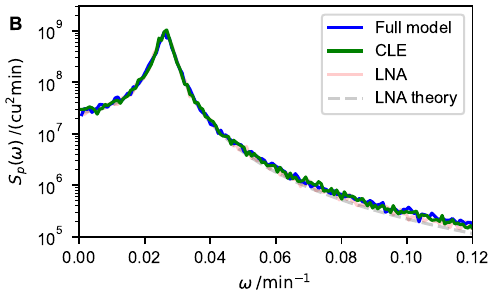}
    \caption{\textbf{Stochastic differential equations accurately describe protein dynamics}. 
    For the following models we show individual trajectories in panel A and power spectra in panel B.
    The full model is a Gillespie simulation of equation~\eqref{eq:full_reactions}, `CLE' stands for chemical Langevin equation (see equation~\eqref{eq:CLE_main}), and LNA stands for `linear noise approximation' (equation~(\ref{eq:lna_main})). `LNA theory' refers to the analytically obtained power spectrum in equation~\eqref{eq:spectra_main}. The parameters used are the same as in figure~\ref{fig:timeseries_noise}, except that here $\Omega=100$ and $\lambda=10\mathrm{min}^{-1}$. All three models exhibit noise-induced oscillations and their power spectra match. Note that $\omega$ represents angular frequencies. To obtain the power spectra, 100 trajectories of each model are generated. They are simulated for $T=10000$min and the first 200 minutes are discarded to account for equilibration. Further details of the computational implementation are available in supplementary section S3.\label{fig:initial_cle_test}}
\end{figure*}

\subsection{Stochastic differential equations accurately describe protein dynamics}\label{sec:sims}
As an initial test of this approximation, we plot individual trajectories of protein concentration $P$ for the full model (equation \eqref{eq:full_reactions}), the chemical Langevin equation \eqref{eq:CLE_main}, and the linear-noise approximation \eqref{eq:lna_main} in figure \ref{fig:initial_cle_test}A. We choose large values of $\Omega$ and $\lambda$ for this test, as that is the regime for which we expect the approximation to work well. Specifically, we use $\Omega = 100$ and $\lambda = 10$min$^{-1}$. All three models exhibit oscillations, indicating that the chemical Langevin equation can capture the effect of noise-induced cycles. Promisingly, all three trajectories appear to have similar qualitative properties, such as the period and amplitude of the oscillations. The potential benefit of using the chemical Langevin equation becomes immediately obvious when comparing computation times: using our implementation, generating trajectories of the chemical Langevin equation for this parameter combination is 463 times faster than generating trajectories of the full model. 

While this visual comparison of trajectories is promising, our next aim must be to evaluate quantitatively how well the two models compare. An ideal measure for this comparison would be one that can characterise the properties of the dynamics visible in figure \ref{fig:initial_cle_test}A. Such a measure is available in the form of the \textit{Fourier power spectrum} of the protein time series. Fourier analysis is a standard method to decompose a time series (such as those in figure \ref{fig:initial_cle_test}A) into contributions from oscillatory signals of varying frequencies. The power spectrum summarises this analysis and captures, for each angular frequency $\omega$, the square of the amplitude, i.e. the strength, of this contribution. Conveniently, the power spectrum of the linear-noise approximation \eqref{eq:lna_main} can be analytically calculated.

We conduct this calculation in the supplementary section S1.6. We find
\BE
S_M(\omega)&=&\frac{(\omega^2+\mu_P^2)\sigma_M^2+[\alpha_M ~f'(P^*)]^2\sigma_P^2}{|\Delta(\omega)|^2},\nonumber \\
S_P(\omega)&=&\frac{\alpha_P^2\sigma_M^2+[\omega^2+\mu_M^2]\sigma_P^2}{|\Delta(\omega)|^2}. \label{eq:spectra_main}
\EE
The denominator in these expressions is
\BE
|\Delta(\omega)|^2&=&\left[\mu_M\mu_P-\alpha_M\alpha_P~f'(P^*)~\cos(\omega\tau)-\omega^2\right]^2 \nonumber \\
&&+\left[\omega(\mu_M+\mu_P)+\alpha_M\alpha_P~ f'(P^*)~\sin(\omega\tau)\right]^2.\nonumber \\
\label{eq:denom_main}
\EE
For the full model and the chemical Langevin equation, the power spectrum can instead be obtained numerically from many repeated simulation runs. 
 
Comparing the protein power spectra for all three models for the same parameters as before (i.e. figure 6e from \cite{manning_et_al2019} and $\lambda=10$min$^{-1}$, $\Omega=100$) reveals that all three models have nearly identical spectra (figure \ref{fig:initial_cle_test}B), highlighting the accuracy of our approximation and demonstrating that our analytical derivations are valid. Hence, the oscillations of the full model in figure \ref{fig:initial_cle_test} are well described by the analytically obtained power spectrum \eqref{eq:spectra_main}.

\begin{figure*}
\includegraphics{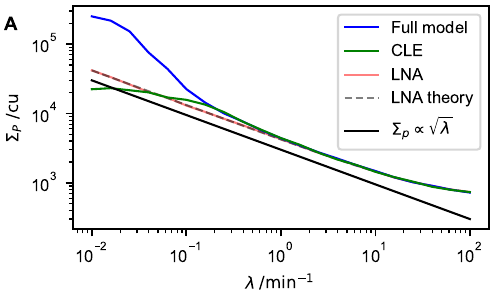} 
\includegraphics{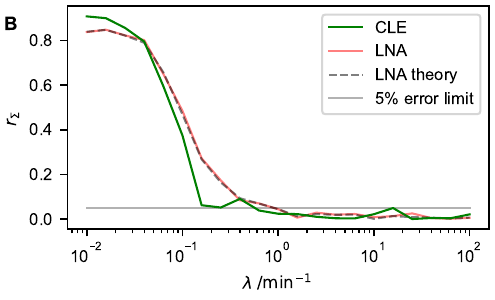}
    \caption{\textbf{Approximation by stochastic differential equations breaks down for small burst frequencies.} (A) Standard deviation of the concentration of protein molecules, $\Sigma_P$, in the stationary state as a function of the bursting parameter, $\lambda$ for multiple models. The labels `Full model', `CLE', `LNA' are defined in figure \ref{fig:initial_cle_test}. Additionally, we include the analytically calculalated standard deviation from equation \eqref{eq:Sigma_p} as `LNA theory'. 
    (B) The relative error $r_\Sigma$ according to equation \eqref{eq:relative_error} for all approximations considered in (A). Parameters are the same as in figure~\ref{fig:timeseries_noise}, with $\Omega=100$. For values of $\lambda$ with $\lambda\le0.1$, two trajectories are created to calculate $\Sigma_P$ and $r_\Sigma$. For values of $\lambda$ with $\lambda >0.1$, 20 trajectories are used. Each trajectory is simulated for $T=10^5$min and the first 2000 minutes are discarded. Further details of the computational implementation are available in supplementary section S3.
    \label{fig:std}}
\end{figure*}

\subsection{Approximation by stochastic differential equations breaks down for small burst frequencies}
Our derivation of the extended chemical Langevin equation \eqref{eq:CLE_main} made the assumption that the bursting parameter, $\lambda$ and the system size, $\Omega$ are large. But what exactly does large mean? Or, in other words, at what values of $\lambda$ or $\Omega$ will the approximation by stochastic differential equations break down? To answer this question, we next seek to computationally estimate this limit by analysing the standard deviation of time series of protein concentration, 
\begin{equation}\label{eq:Sigma_P_def}
\Sigma_P = \sqrt{\avg{P^2}_t-\avg{P}_t^2},
\end{equation}
where $\avg{\cdots}_t$ stands for a time average in the stationary state, i.e. after long waiting times. This quantity is a measure of the amplitude of noise-induced oscillations about a deterministic fixed point. Where the power spectrum is analytically available, the standard deviation can be calculated from the power spectrum $S_P$ via the relation 
\begin{equation}\label{eq:Sigma_p}
\Sigma_P^2=\frac{1}{2\pi}\int_{-\infty}^\infty d\omega S_P(\omega).
\end{equation}
A derivation of this relationship is provided in the supplementary section S1.7.
We can use equation \eqref{eq:spectra_main} to analytically predict $\Sigma_P$ from the linear-noise approximation.

We analyse how the standard deviation in protein time series of each model, i.e. the full model \eqref{eq:full_reactions}, the chemical Langevin equation \eqref{eq:CLE_main}, and the linear-noise approximation \eqref{eq:lna_main} depends on the bursting parameter $\lambda$ for a fixed system size of $\Omega=100$ (figure~\ref{fig:std}A). The theory derived within the linear-noise approximation captures the standard deviation $\Sigma_P$ of the full model quantiatively for sufficiently large burst frequencies $\lambda$. When bursting is slow on the other hand, significant deviations between the linear-noise approximation and simulations of the full model are observed. 

To be able to accurately identify the point of deviation between the models, we define the relative error to the full model
\begin{equation}
r_\Sigma(\lambda) = |\Sigma_{P}(\lambda) - \Sigma_{P,F}(\lambda)|/\Sigma_{P,F}(\lambda)\label{eq:relative_error}
\end{equation}
where $\Sigma_P$ indicates the protein standard deviation of the model in question (the chemical Langevin equation \eqref{eq:CLE_main}, the linear-noise approximation \eqref{eq:lna_main} or the theoretical result using equations \eqref{eq:Sigma_p} and \eqref{eq:spectra_main}), and $\Sigma_{P,F}$ refers to the protein standard deviation of the full model \eqref{eq:full_reactions}.

The linear-noise approximation becomes accurate to at most a $5\%$ error in this example roughly when bursts occur at least once per minute ($\lambda\approx 1{\rm min}^{-1}$) (figure \ref{fig:std}B). Even for much slower burst frequencies, the approximation produces reasonable ball park estimates for the magnitude of fluctuations (and hence for the amplitude of noise-induced oscillations). The relative error between the linear-noise approximation and the full model remains below one for burst frequencies on a time scale of $1/\lambda\approx 100{\rm min}$. That is, the linear-noise approximation produces estimates that are within a factor of two of the true standard deviation for burst frequencies with time scales of approximately an hour. 

Figure~\ref{fig:std}A also shows that the standard deviation $\Sigma_P$ reduces for increased burst frequencies $\lambda$. This is because the strength of bursting noise scales as $\lambda^{-1/2}$ (to leading order). Thus, for high values of $\lambda$, the promoter's ON/OFF dynamics average out. Fluctuations induced by bursting then become less pronounced in the dynamics of the gene regulatory network. We would not expect $\Sigma_P$ to approach zero in the limit of infinitely fast bursting ($\lambda\to \infty$). This is because the effects of intrinsic noise remain in this limit, so we expect oscillations purely induced by copy-number noise, as observed for example in figure~\ref{fig:timeseries_noise} for the model with only copy-number noise.

The system-size parameter is held fixed at $\Omega=100$ in figure~\ref{fig:std}. For very slow bursting dynamics (small values of $\lambda$) we then expect bursting noise to be much stronger than copy number noise. Oscillations are then predominantly induced by bursting noise. Within the linear noise approximation, their amplitude can be expected to scale as $1/\sqrt{\lambda}$, and the contributions containing $1/\Omega$ in equations~(\ref{eq:sigma}) and (\ref{eq:spectra_main}) become irrelevant. This is indeed what we see in figure~\ref{fig:std}A.

\begin{figure}[t!]
    \centering
    \includegraphics{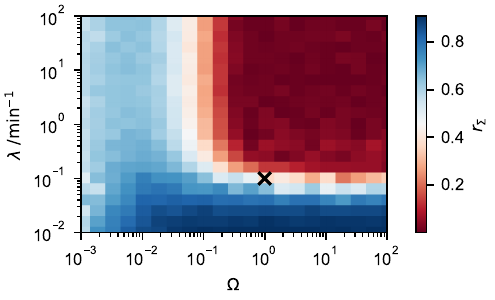}
    \caption{\textbf{Relative error $r_\Sigma$ of standard deviation of the chemical Langevin equation (\ref{eq:CLE_main}) compared to the full model in equation~(\ref{eq:reactions})}. The relative error is calculated using equation~\eqref{eq:relative_error}. The black cross represents the grid point $\lambda=0.1$min$^{-1}$ and $\Omega=1$. Parameters are the same as in Figure 2. For gridpoints where $\lambda\le0.1$min$^{-1}$, two trajectories are created to calculate $r_\Sigma$. For gridpoints where $\lambda >0.1$min$^{-1}$, 20 trajectories are created. Each trajectory is simulated for $10^5$min, and the first 2000 minutes are discarded. Details of the computational implementation are available in supplementary section S3.}
    \label{fig:heatmap}
\end{figure}

\subsection{Suitability of our approximation is application dependent}\label{sec:realism}
Our analysis of the breakdown of the approximation in figure \ref{fig:std} assumed that $\Omega=100$, i.e. we only identified a critical value of $\lambda$. However, our derivation of the chemical Langevin equation \eqref{eq:CLE_main} required us to assume that both $\Omega$ and $\lambda$ are large. Hence, we would expect the accuracy of the extended chemical Langevin equation to depend on both $\Omega$ and $\lambda$, with the possibility that, for each $\Omega$, different values of $\lambda$ are required for accuracy of the approximation. 

We analyse the joint dependence of the accuracy of the chemical Langevin equation \eqref{eq:CLE_main} on $\lambda$ and $\Omega$ in figure \ref{fig:heatmap}, again focussing on our measure of relative error \eqref{eq:relative_error}. We find that the predictions from the chemical Langevin equation reproduce the standard deviation of protein fluctuations up to an error of at most $5\%$ when $\Omega\gtrsim 0.3$ and $\lambda \gtrsim 0.5 {\rm min}^{-1}$ (figure~\ref{fig:heatmap}).
In the figure, all other parameters are as in figures~\ref{fig:timeseries_noise}-\ref{fig:std}, matching those in  \cite{manning_et_al2019}.

While our choice of $P_0=24201$cu together with a value of $\Omega=1$ is consistent with experiments \cite{manning_et_al2019}, the requirement $\lambda\gtrsim 0.5{\rm min}^{-1}$ indicates that the bursts would have to occur once every two minutes or faster for the chemical Langevin equation to become accurate to an error of less than $5\%$.
That leaves us to ask in which regime experimentally measured value for the burst parameter $\lambda$ lie. In our model \eqref{eq:CLE_main}, $\lambda$ represents the rate at which the promoter switches from the OFF to the ON state. This rate varies widely between individual genes, and between organisms. For example, live-imaging in yeast suggested $\lambda$ of 1-4min$^{-1}$ \cite{fu}, whereas measurements on a range of bacterial genes found 0.3min$^{-1}$ \cite{so}. Measurements in the fruit fly \textit{Drosophila} indicate values of around 0.15min$^{-1}$ by \cite{zoller} and up to 2min$^{-1}$ by \cite{pimmett}. Bursting in mammalian cells is considered to be slower than in other organisms, with values of around 0.01min$^{-1}$ reported for a range of genes by \cite{suter}. Yet, for HIV-1 RNA in HeLa cells, bursting frequencies of up to 0.6min$^{-1}$ have been observed \cite{tantale}.  

Considering these experimental measurements, we investigate the representative value $\Omega=1$, $\lambda=0.1{\rm min}^{-1}$ in figure \ref{fig:heatmap}. At this parameter combination, protein fluctuations in the extended chemical Langevin model deviate from that in the full model by approximately $37\%$ (figure \ref{fig:heatmap}). For larger $\lambda$, the approximation becomes more accurate, for smaller $\lambda$, the approximation eventually breaks down. This illustrates that the applicability of our method will depend on the biological scenario: in some cases, an approximation using our extended chemical Langevin equation will work well. In other situations it will be more suitable to use a piecewise continuous chemical Langevin equation, i.e. a model following Gillespie's derivation to approximate copy-number noise \cite{gillespieCLE}, but simulating promoter switching explicitly, such as in equation \eqref{eq:pdmp}. For small system sizes $\Omega$ and small $\lambda$, it will be necessary to use the full model, equation \eqref{eq:full_reactions}. This is a workable approach, as stochastic simulations using the Gillespie algorithm will be fast in this scenario. 

The breaking point of the approximation in our model is at around $\lambda = 0.5$min$^{-1}$. This breaking point may be different for other transcription factor networks: the timescale of fluctuations of gene regulatory networks is typically set by degradation rates \change{(\cite{alon}, chapters 1-3)} and one may expect that the applicability of our Gaussian approximation depends on how the bursting parameter $\lambda$ relates to degradation rate of the mRNA that is being produced. For more long-lived mRNAs, such as those involved in the circadian clock, slower bursting frequencies may be permissive in our approximation. In a given application, it will hence be beneficial to test the applicability of the chemical Langevin approximation through comparison to test simulations of the corresponding full, unapproximated model. 

\change{
We expect that the requirement of sufficiently large $\Omega$ is met for a large number of genes. Specifically, we expect $\Omega$ to be sufficiently large if the number of protein molecules per nucleus exceeds a few hundred to a few thousand. Our figure \ref{fig:heatmap} illustrates this, as the relative error falls below 5\%  as $\Omega>0.5$, which would correspond to around 20000 protein molecules per nucleus. More than half of all quantified genes in a global screen conducted by \cite{schwanhausser} fall within that range.
}

The chemical Langevin equation we derived assumes Gaussian increments in mRNA and protein copy number between consecutive time intervals. In cases where the approximation breaks down when the bursting parameter $\lambda$ is low, we expect increments to be \textit{non-Gaussian}, suggesting that slow transcriptional bursting is an inherently non-Gaussian effect. This is consistent with experimental observations that mRNA distributions can be non-Gaussian and non-Poissonian \cite{munsky}.

\subsection{Stochastic differential equation can approximate other gene regulatory networks}
In equation \eqref{eq:full_reactions}, we chose the rate $G_{ON}\rightarrow G_{OFF}$ to depend on the current protein concentration $P$. This follows a model for repressive transcription factors initially proposed by \cite{alon} \change{(see appendix A of the reference)}, and is designed to generate the typical Hill function \eqref{eq:hill} in the stationary limit. However, alternative functional implementations of the action of a transcription factor are possible. For example, multiple recent studies suggest that transcriptional repressors affect the rate of $G_{OFF}\rightarrow G_{ON}$ transitions, thus modulating the waiting times in the OFF period (e.g. \cite{zhao}). Our framework of chemical Langevin equations to approximate the effect of promoter switching is suitable for such alternative parameterisations, and we illustrate this in supplementary section S2 and supplementary figures SF1 and SF2.

\begin{figure}
    \centering
   \includegraphics{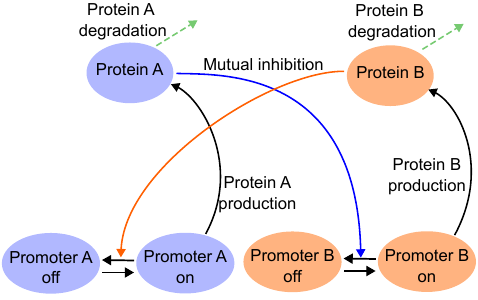}
    \caption{\textbf{Schematic of the toggle switch model.} The model describes two protein concentrations $A$ and $B$, and the effects of separate promoters for each. These can be in an active or an inactive state (on/off). Both proteins can be degraded. The promoter of each protein can be inactivated by the presence of the respective other protein.}
    \label{fig:toggle_schematic}
\end{figure}

Similarly, our method is applicable to other gene regulatory networks. A seminal example of such a gene network is the bistable toggle switch, which has been successfully engineered into EColi bacteria \cite{gardner}. The standard model for this switch considers two protein concentrations without representing mRNAs or delays. The two protein concentrations mutually repress each other and can each be degraded. This model can be extended to account for multiple promoter states (figure \ref{fig:toggle_schematic}), see Jia et al. \cite{holimap}. Choosing a promoter parameterisation for both components of the switch inspired by equation \eqref{eq:full_reactions}, we arrive at the chemical reaction system
\BE
G_{A,\text{ON}} &\stackrel{\lambda (B/P_0)^h}{\longrightarrow}& G_{A,\text{OFF}}, \nonumber \\
G_{A,\text{OFF}} &\stackrel{\lambda }{\longrightarrow}& G_{A,\text{ON}}, \nonumber \\
G_{B,\text{ON}} &\stackrel{\lambda (A/P_0)^h}{\longrightarrow}& G_{B,\text{OFF}}, \nonumber \\
G_{B,\text{OFF}} &\stackrel{\lambda }{\longrightarrow}& G_{B,\text{ON}}, \nonumber \\
A&\stackrel{\mu}{\longrightarrow}& \varnothing, \nonumber \\
B&\stackrel{\mu}{\longrightarrow}& \varnothing, \nonumber \\
\varnothing &\stackrel{\alpha \sigma_A}{\longrightarrow}& A, \nonumber \\  
\varnothing &\stackrel{\alpha \sigma_B}{\longrightarrow}& B, \nonumber \\  
\label{eq:toggle_reactions}
\EE
where $G_{\cdot,\text{ON}}$ represents promoter ON states for proteins A or B, respectively, and $G_{\cdot,\text{OFF}}$ represents the promoter OFF states.
The bursting parameter $\lambda$ regulates the burst frequency and duration in the ON states of both promoters. Each promoter experiences the repression by the respective other protein, with the same repression threshold $P_0$ and the same Hill coefficient $h$.
Both proteins are degraded with degradation rate $\mu$, and in the respective ON states of the promoter the proteins are produced by the transcription rate $\alpha$. 
The auxiliary variable $\sigma_A$ is equal to one if the A promoter is in the ON State $G_{A,\textbf{ON}}$ and equal to zero if the promoter is in the OFF state $G_{A,\textbf{OFF}}$. Similarly, $\sigma_B$ takes the values 1 and 0 depending on the state of the B promoter. We have chosen both proteins to share the same parameters for production, degradation and repression for simplicity. The concentrations $A$ and $B$, are given by $n_A/V$ and $n_B/V$, respectively, where $n_A$ and $n_B$ are the number of $A$ and $B$ molecules, and $V$ refers to the nuclear volume. For convenience, we again write $V=\Omega V_0$, with $V_0$ determining the volume of a reference nucleus. Concentrations are again measured in \textit{concentration units} cu, referring to one molecule per volume $V_0$.

\begin{figure*}
    \centering
   \includegraphics{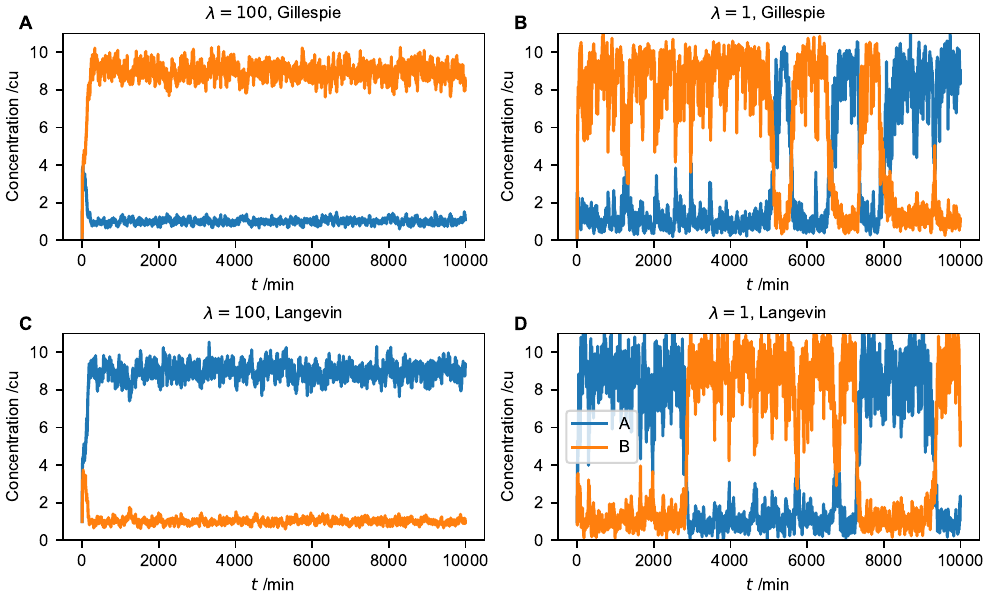}
    \caption{\textbf{Noise induced transitions in the toggle-switch model.} (A) A representative simulation of the toggle switch model in \eqref{eq:toggle_reactions} using the parameter combination $P_0 = 3$cu, $h = 2$, $\mu = 0.1$min$^{-1}$, $\alpha = 1$cu/min, $\lambda = 100$min$^{-1}$ and $\Omega = 100$, simulated using the Gillespie stochastic simulation algorithm \cite{gillespiessa}. (B) A representative simulation of the toggle switch model \eqref{eq:toggle_reactions} using the same parameters as in (A), except that now $\lambda = 1$min$^{-1}$ (C) A representative simulation of chemical Langevin equation for the toggle switch model, equation \eqref{eq:CLE_toggle}, using the same parameters as in (A), simulated using an Euler-Maruyama scheme and a timestep of 0.01min. The dynamics are similar to (A), except that the stationary concentration profiles of $A$ and $B$ are swapped. (D) A representative simulation of chemical Langevin equation for the toggle switch model, equation \eqref{eq:CLE_toggle}, using the same parameters as in (B).}
    \label{fig:toggle_illustration}
\end{figure*}

Intriguingly, this system can exhibit transitions between two steady states that are enabled by bursting noise. Specifically, for the parameter combination of $P_0 = 3$cu, $h = 2$, $\mu = 0.1$min$^{-1}$, $\alpha = 1$cu/min and $\Omega = 100$, we see bistability for $\lambda = 100$min$^{-1}$. Starting from an initial condition of the concentrations $A=1$cu and $B = 1$cu we find that the concentration of $B$ molecules approaches a high steady state, fluctuating around the value of approximately 9cu (figure \ref{fig:toggle_illustration}A), whereas the concentration of A fluctuations around a lower value of approximately 1cu. Swapped stationary concentrations for $A$ and $B$ occur in different realisations of the process. Using this same parameter combination, but with $\lambda = 1$min$^{-1}$, stochastic transitions can be observed, where each of the concentrations $A$ and $B$ is low during some periods and high during others (figure \ref{fig:toggle_illustration}B). 
The parameter combinations used in figure \ref{fig:toggle_illustration}A and B thus identify a regime in which promoter bursting is functionally important. Therefore we test our approximation by chemical Langevin equations for this parameter combination.

The chemical Langevin approximation of this system emerges as
\BE\label{eq:CLE_toggle}
\dot A &=& f(B)-\mu A +\xi_A(t), \nonumber \\
\dot B &=& f(A)-\mu B +\xi_B(t), 
\EE
with Gaussian white noise variables $\xi_A(t)$ and $\xi_B(t)$, each of mean zero, and with 
\BE
\avg{\xi_A(t)\xi_A(t')}&=&\frac{\alpha^2}{\lambda}\frac{2[B/P_0]^h}{\big(1+[B/P_0]^h\big)^3} \delta(t-t') \nonumber\\
&&\left.\hspace{-5em} + \frac{1}{\Omega} \big[f(B) + \mu A\big]\right.\delta(t-t'),\nonumber \\
\avg{\xi_B(t)\xi_B(t')}&=&\frac{\alpha^2}{\lambda}\frac{2[A/P_0]^h}{\big(1+[A/P_0]^h\big)^3} \delta(t-t') \nonumber\\
&&\left.\hspace{-5em} + \frac{1}{\Omega} \big[f(A) + \mu B\big]\right.\delta(t-t'), \nonumber \\
\avg{\xi_A(t)\xi_B(t')}&=&0. \label{eq:noise_cov_toggle}
\EE
Indeed, we find that the chemical Langevin equations approximate the dynamics of the toggle switch well. Protein concentrations simulated by the chemical Langevin equation \eqref{eq:CLE_toggle} fluctuate on a timescale comparable to those simulated by the Gillespie algorithm, and have comparable amplitude (compare figures \ref{fig:toggle_illustration}C and \ref{fig:toggle_illustration}D with \ref{fig:toggle_illustration}A and \ref{fig:toggle_illustration}B, respectively). These simulations by the chemical Langevin equation also capture the noise-induced transitions between the two states of the model when comparing the cases $\lambda = 1$min$^{-1}$ and $\lambda = 100$min$^{-1}$ (figures \ref{fig:toggle_illustration}C and \ref{fig:toggle_illustration}D).

\begin{figure*}
    \centering
   \includegraphics{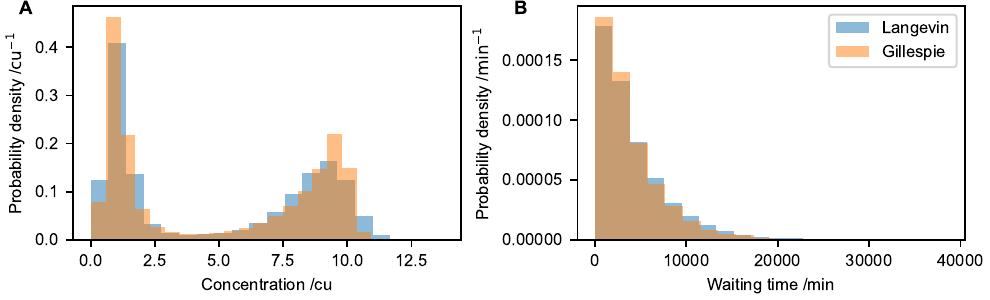}
    \caption{\textbf{Chemical Langevin equations approximate the toggle switch dynamics well.} (A) Stationary distribution of the concentration of A molecules for simulations of the Gillespie stochastic simulation algorithm in equation \eqref{eq:toggle_reactions}, and the chemical Langevin equation \eqref{eq:CLE_toggle}. (B) Distribution of waiting times to switch from the configuration ($A$ high, $B$ low), to ($A$ low, $B$ high), for both models. For both panels, the same parameter combination as in figure \ref{fig:toggle_illustration}B was used. Both panels analyse simulation output from one simulation of each model, that was run for a duration of $T=10^7$min. The chemical Langevin equation was simulated using an Euler-Maruyama scheme with a timestep of 1min. To calculate waiting times, each timeseries was smoothened by a sliding average of 1000min. Then, the length of any time window was recorded in which the smoothened timeseries of the A concentration was greater than 4cu, and where the smoothened concentration of B was less than 4cu. The histogram is generated from all waiting times recorded in this way. }
    \label{fig:toggle_comparison}
\end{figure*}

Importantly, the stationary distribution obtained from Gillespie simulations of the full model \eqref{eq:toggle_reactions} is well approximated by the chemical Langevin equations \eqref{eq:CLE_toggle} (figure \ref{fig:toggle_comparison}A). The means and standard deviations of both distributions are within <0.5\% of each other. Similarly, the distribution of waiting times in each of the stable states is well approximated by the chemical Langevin equation (figure \ref{fig:toggle_comparison}B). The means and standard deviations of the waiting time distribution sampled from the chemical Langevin equation differ by less than 10\% from the distribution generated using the Gillespie stochastic simulation algorithm (for approximately 11,000 sampled waiting times).

In summary our chemical Langevin approximation is applicable and accurate for other gene regulatory networks, following the theory provided in supplementary section S1.

\section{Summary and discussion}\label{sec:summary}
The importance of noise in gene regulatory networks is widely recognised \cite{raj,munsky,porello}. Here, we investigated how noise due to transcriptional bursting affects the dynamics of the broadly studied auto-negative feedback motif and found that bursting can induce oscillations when they are not otherwise present, and enhance oscillations that may already be present in the system. We provided computationally efficient approximations of the system dynamics and illustrated the utility of our approximation through analytical derivations of the power spectrum of the observed oscillations. Simulations of the chemical Langevin equation are multiple orders of magnitude faster than those of the full system. Our calculations confirm mathematically that transcriptional bursting alone can induce oscillations, and that transcriptional bursting can enhance the amplitude of oscillations if these are already present due to copy number noise. We investigated the breaking point of our extended chemical Langevin equations and related it to biological measurements of transcriptional burst frequencies and showed how our theory can be extended to other gene regulatory networks.  

Our models considered a single active promoter for each gene. Our approach is nonetheless applicable to systems with more active promoters, i.e. genes with more than one active allele in haploid or even multiploid cells \cite{strasser,lin2,Luo2023}. This would be implemented by assuming more transcriptional states, representing the different combinations of promotor activity on each allele. We further highlight that the observed oscillations in our model are only visible in the stationary regime, i.e. when a gene is constitutively expressed, and cannot be used to describe genes that recently turned on, for example.

Varying definitions of burstiness exist in the literature. The concept was originally introduced to describe mRNA production that is discontinuous in time \cite{chubb,porello}). This is the phenomenon we seek to study here. Other authors exclusively consider the gene to undergo transcriptional bursting if the OFF-periods of the promoter are much longer than the ON-periods \cite{larsson, paulsson,jia2017}. Our approximations are applicable under both of these definitions of burstiness, i.e. they can describe systems where OFF-periods are much longer than ON-periods, but this is not a necessary restriction for the validity of our methods. We are instead focused on the consequences of the random switching between the OFF and the ON states of the promoter, rather than their relative duration. 

In the language of transcriptional bursting, our bursting parameter $\lambda$ affects the duration of ON-periods and of OFF-periods of the promoter, and therefore the bursting initiation rate as well as the burst size, i.e. the number of transcripts per transcriptional burst. This dependency is chosen such that, when varying the bursting parameter $\lambda$, we vary the timescale of bursting, but not the expected relative durations of ON- and OFF periods. This joint dependency of both rates on $\lambda$ is necessary to ensure that our changes in promoter dynamics do not affect the average transcription rate of the promoter, which is known to impact the oscillatory properties of the model \cite{monk}.

Following these design choices, the limits  $\Omega\rightarrow \infty$ and $\lambda\rightarrow\infty$ allow us to construct situations in which either copy-number noise or promotor-bursting noise are `turned off'. Being able to do this is a distinct advantage of the model, as it means we can investigate the consequences of each type of noise individually. Identifying protocols to perform analogous investigations experimentally would be challenging, and may not be possible with existing technologies. For example, to change the system size parameter, the experiment of interest would require (i) an increased production of mRNA, (ii) a decreased sensitivity of the promoter to the number of protein molecules, and (iii) maintaining an identical burst frequency while doing so. Increased production of mRNA (i) may be achieved through upstream regulation by a relevant transcription factor. However that may also affect bursting kinetics. Changing the sensitivity to protein may be achieved through binding of protein molecules via dimerisation or modifying concentrations of a transcriptional co-regulator. Both of these would need to be done in a way that does not affect the bursting kinetics. Similarly, changing $\lambda$ would require changing the bursting kinetics in a way that affects both the burst frequency as well as the burst size, which would possibly require a modification of the promoter or enhancer sequences, as well as a change in upstream regulation. In summary, the model allows us to achieve understanding that would be difficult to obtain by experiments alone.

Our investigation continues a long line of research on the delayed auto-negative feedback motif \cite{monk,lewis,momiji,philips,barrio,galla2009,brett,burton,manning_et_al2019,alon,hornos,grima,kumar,jia2019,jia2020,ramos,jia2020a,wu}. Among this work, \cite{momiji} pointed out that the model presented in equation \eqref{eq:monk} is insufficient to explain the full amplitude of biologically observed oscillations, and they hence argued that alterations to the model, such as the consideration of protein dimerisation or periodic forcing, are necessary to alleviate this discrepancy. Our results suggest that instead the presence of transcriptional bursting may be sufficient to explain an increase in oscillation amplitude, a result in line with previous findings on a bursting model with negative feedback \cite{jia2024}. We expect the auto-negative feedback motif to remain relevant as auto-repression is abundant in gene regulatory networks, suggesting the existence of yet unknown oscillators \cite{minchington}.

Our work contributes to existing efforts to efficiently simulate the dynamics of gene regulatory networks, such as the methods presented by \cite{hettich}. Simulating chemical Langevin equations is generally much less demanding in terms of computing time than the simulation of the full reaction system from which the CLE is derived. Where our approximation is applicable, this means that equations such as (\ref{eq:CLE_main}) are a viable approach to studying the role of transcriptional bursting in natural gene regulatory networks. For example, high-throughput simulations of the chemical Langevin model can be used to infer model parameters from experimental data \cite{burton,browning}.

When deriving our extended chemical Langevin equation in supplementary section S1, we introduce the promoter states as \textit{environmental states} in a chemical reaction system. The response of chemical reaction systems to rapidly switching environments in this way was previously derived in \cite{hufton_pre_1,hufton_pre_2}. This work also included levels of approximation for large copy numbers or fast environmental switching rates, respectively. In these studies, the authors used a Kramers-Moyal expansion to arrive at an extended chemical Langevin equation. Our analysis expands this previous work by providing an alternative derivation of the extended Chemical Langevin equation in the style of \cite{gillespieCLE}. This derivation is advantageous to us over the Kramers-Moyal expansion, as it allows a natural inclusion of delayed terms. Without this inclusion of delayed terms, our discussion of the delayed auto-negative feedback motif would not have been possible.

Our work opens up possibilities in a wider range of areas. Copy number fluctuations are known to generate oscillations not only in gene regulatory circuits but also in other biochemical systems, in models of epidemic spread and in evolutionary game theory \cite{alan2,alonso,bladon}. Such stochasticity , sometimes called `demographic noise' in other areas, can also provoke spatial patterns or waves \cite{biancalani,karig,biancalani2}. It is conceivable that these phenomena can also be induced by extrinsic noise, or be enhanced by it. Further work is required to investigate this possibility.

In the context of gene regulatory networks, we hope that our discussion of the extended chemical Langevin equation, as well as the accompanying approximations for large or finite burst frequencies or system sizes will facilitate the wider study of gene regulation and cell fate.


\section{Author contributions}
JK: conceptualisation, writing - original draft, writing - review and editing, supervision, investigation (derivation of chemical Langevin equations), methodology, software (edited code and finalised figures for main model, wrote code and generated figures for the alternative models); AM: visualisation, writing - review and editing, investigation (parameter variations), methodology, software (wrote draft code for main model);  TG: conceptualisation, writing - original draft, writing - review and editing, supervision.

\section{Funding statement}
This work was partially supported by the María de Maeztu project CEX2021-001164-M funded by the  MICIU/AEI/10.13039/501100011033.

\section{Data accessibility}
All code used in this study is available via the repository \url{https://zenodo.org/records/15046770}.

\bibliography{references}

\onecolumn
\clearpage

\setcounter{page}{1}

\title{}

\begin{center}
{
\noindent
\Large{\textbf{Supplementary material to the article `Efficient approximations of transcriptional bursting effects on the dynamics of a gene regulatory network' in the Journal of the Royal Society Interface}}
}

\bigskip
\noindent Jochen Kursawe${}^{1}$, Antoine Moneyron${}^{2}$, 
Tobias Galla${}^{3}$ \\~\\
 
\end{center}

\noindent
${}^{1}$ School of Mathematics and Statistics, University of St Andrews, North Haugh, St Andrews, KY16 9SS, United Kingdom

\noindent
${}^{2}$ Inria de l'Université de Rennes, Campus de Beaulieu, 263 Av. Général Leclerc, 35042 Rennes, France

 \noindent
${}^{3}$ Instituto de F\'isica Interdisciplinar y Sistemas Complejos, IFISC (CSIC-UIB), Campus Universitat Illes Balears, E-07122 Palma de Mallorca, Spain
\\
\vspace{0.5cm}

\setcounter{figure}{0}
\setcounter{equation}{0}
\setcounter{section}{0}
\setcounter{page}{1}
\renewcommand{\thesection}{S\arabic{section}} 
\renewcommand{\theequation}{SE\arabic{equation}}
\renewcommand{\thefigure}{SF\arabic{figure}}
\renewcommand{\thepage}{SP\arabic{page}}

\section{Derivation of a chemical Langevin equation for fast promoter bursting}\label{sec:sm_cle}
In this section we derive the chemical Langevin description for a gene regulatory network that takes bursting noise into account. To do so, we extend the classical derivation by Gillespie \cite{gillespieCLE}. Our approximation is valid for systems with large but finite copy numbers, and provided that the promoter switches between states on a time scale that is much faster than that of the remaining system. Nonetheless, we do not require infinitely fast promoter bursting. Instead, our approximation captures leading order corrections to the full adiabatic elimination. It also describes fluctuations that arise due to finite copy numbers to leading order.

In our derivation, we consider promoter states as \textit{environmental states} of a chemical reaction system. These environmental states are defined such that they affect the reaction propensities of the rest of the system. 
In this supplementary material, we first present the approximation through a chemical Langevin equation for a general reaction model that includes delayed reactions and switching between environmental states. Then, we discuss the case where the environment can stochastically switch between two states, i.e. where the environment represents a telegraph model of transcription. Finally, we discuss the special case of the auto-negative feedback motif described in the main paper.

\subsection{The promoter defines \textit{environmental states} in a system of chemical reactions}
Consider a chemical reaction system with molecules of types $i=1,\dots,S$, and with reactions $r=1,\dots,R$. We write $n_i(t)$ for the number of particles of type $i$ at time $t$, and thus $\bm{n}(t) = [n_1(t),...,n_S(t)]$ describes the chemical state of the reaction system at time $t$. We additionally assume that the system may operate in one of $M$ \textit{environmental states} $\sigma=1,\dots,M$. The environmental states are defined to affect the reaction propensities, which allows us to model chemical reaction systems that can change their reaction rates and structure under different `environmental' conditions. The environmental states may themselves change in time via deterministic or probabilistic rules. In our specific model in equation (1) of the main manuscript, there are two environmental states, which correspond to the two states that the promoter can assume. 

We introduce the notation $g_{r,\sigma}$ for the propensity of reaction $r$ to trigger at time $t$, given that the environmental state is $\sigma$. If the system is in state $\sigma$ the probability that reaction $r$ triggers in the next infinitesimal time element $dt$ is then $g_{r,\sigma}(t)dt$. The propensities $g_{r,\sigma}(t)$ only depend on the chemical state of the system, $\bn(t)$, and on the environmental state, $\sigma(t)$, at time $t$. 

The dynamics in a short time interval can then be written as
\be
\label{eq:gillespie_ansatz}
n_i(t+\Delta t)=n_i(t)+\sum_r \nu_{i,r} k_r(t)
\ee
where the $\nu_{i,r}$ are stochiometric coefficients (the number of particles of type $i$ changes by $\nu_{i,r}$ in a reaction of type $r$), and where $k_j(t)$ is the number of reactions of type $r$ firing in the time interval from $t$ to $t+\Delta t$. 



\medskip

The $k_r(t)$ in equation \eqref{eq:gillespie_ansatz} are random variables, and their distribution depends on the reaction propensities. Equation \eqref{eq:gillespie_ansatz} is an approximation-free statement: if $k_r(t)$ is known, it is true for any combination of chemical and environmental states at all times $t$. Our aim is to make simplifying assumptions that allow us to approximate this ansatz \eqref{eq:gillespie_ansatz} using a stochastic differential equation.

To do so, we follow the steps originally proposed by Gillespie \cite{gillespieCLE} and assume that $\Delta t$ is sufficiently small such that that the $\sigma$-dependent reaction propensities $g_{r,\sigma}$ are all constant in the time interval from $t$ to $t+\Delta t$ for each fixed value of $\sigma$. We add to Gillespie's derivation, by assuming that the propensities by which the environment changes between its $M$ states are also constant throughout $[t, t+\Delta t]$.

Under these assumptions, the $k_r(t)$ are mixtures of Poisson random variables. In particular, if the system switches between multiple environmental states in the time interval $[t,t+\Delta t]$, then within each time period of constant environmental state the number of each reaction in this period is a Poisson random variable. Consequently, for a given trajectory of the environment in $[t, t+\Delta t]$, $k_r(t)$ is a sum of independent Poisson random variables, which itself is a Poisson random variable. If the environment can change under probabilistic rules, then $k_r(t)$ will therefore be a mixture of Poisson random variables.

We proceed to make a second assumption inspired by \cite{gillespieCLE}, namely that $\Delta t$ is sufficiently large such that the $k_r(t)$ are further approximated by Gaussian random variables. This assumption invokes the Central Limit theorem and therefore relies on the previous assumption that the propensities within the chemical reaction system, as well as the propensities of the environment to change, are constant throughout $[t,t+\Delta t]$. Our second assumption further includes that many environmental changes occur in the time interval $[t,t+\Delta t]$. 

The next step in our deriviation is to calculate the means and variances of these Gaussian random numbers.

\subsection{Calculation of mean and variance of number of reactions in one time step} 

\subsubsection{Mean}\label{sec:mean}
To calculate the mean number of each reaction in the time step $\Delta t$, we first consider a case where environmental changes happen at fixed points in time, i.e. we condition on one trajectory of the environment. Within the time interval $[t,t+\Delta t]$ the environment then spends a total time of $\Delta t_1$ in environment $1$, total time $\Delta t_2$ in environment $2$ and so on, with $\Delta t=\sum_\sigma \Delta t_\sigma$. Since the $g_{r,\sigma}$ do not contain any explicit time dependence, the quantity $k_r(t)$ in this scenario is a sum of independent Poisson random variables
\be
k_r(t)=\sum_\sigma k_{r,\sigma}(t),
\ee
where $k_{r,\sigma}(t)$ has mean $g_{r,\sigma}(t)\Delta t_\sigma$.
The sum of independent Poisson random variables is a Poisson variable, with the rate parameter given by the sum of the individual rate parameters. Hence $k_r(t)$ is a Poissonian random variable with mean $\sum_\sigma g_{r,\sigma}(t)\Delta t_\sigma$. Defining
\be\label{eq:gbardef}
\overline{g}_r(t)\equiv \sum_\sigma \frac{\Delta t_\sigma}{\Delta t} g_{r,\sigma}(t),
\ee
we can identify $k_r(t)$ as a Poissonian random variable with mean $\overline g_r \Delta t$, where $g_r$ emerges as the average reaction propensity of reaction channel $r$.

These observations are true if the environment only undergoes deterministic changes. If the environment undergoes probabilistic changes, then the same reasoning holds, except that the quantity $\overline{g}_r$ will itself will be a random variable, due to the randomness of the time intervals $\Delta t_\sigma$ spent in the different environmental states. We write $P(\overline{g}_r)$ for the probability density of $\overline{g}_r$. The overall average number of reactions of type $r$ in the time interval from $t$ to $t+\Delta t$, $\avg{k_r(t)}$ is then given by
\be
\avg{k_r(t)}=\int k_r P(k_r)\mathrm{d}k_r = \iint k_r P(k_r| \overline{g}_r) P(\overline g_r)\mathrm{d}\overline g_r\mathrm{d}k_r.\label{eq:average_k_r}
\ee
In the first integral we have simply used the mathematical definition of the mean, using $P(k_r)$ to denote the probability density to observe $k_r$ reaction events of type $r$. In the second integral, we used the law of total probability to link $P(k_r)$ to $\overline{g}_r$ within the conditional probability $P(k_r| \overline{g}_r)$ to observe $k_r$ reaction events of type $r$ given that the sample path of the environment leads to an average reaction propensity of $\overline{g}_r$. Importantly, we have that  
\be
\int k_r P(k_r| \overline{g}_r) \mathrm{d}k_r = \Delta t\, \overline{g}_r
\ee
since the $P(k_r| \overline{g}_r)$ are Poisson distributions with mean $\overline{g}_r\Delta t$, by our reasoning above. Therefore, equation \eqref{eq:average_k_r} simplifies to
\be
\avg{k_r(t)}=\Delta t\int d\overline g_r  ~\overline g_r P(\overline g_r) =:\Delta tE_{\overline g_r}(\overline{g}_r).
\ee
If the environment is in equilibrium we can assume that $\int d\overline g_r  ~\overline g_r P(\overline g_r)$ is the average rate of reaction $r$ in the stationary state of the environment.

\subsubsection{Variance}


Having identified the mean of the increments in equation \eqref{eq:gillespie_ansatz}, we now calculate the variance of the increments. To do so, we apply the law of total variance
\be\label{eq:tot_var}
\mbox{var}(k_r)= E_{\overline{g}_r}[\mbox{var}(k_r|\overline{g}_r)]+\mbox{var}_{\overline g_r}[ E(k_r|\overline g_r)],
\ee
where $E_{\overline g_r}(\cdots)$ stands for an average over all possible average propensities $\overline g_r$, and $\mbox{var}_{\overline g_r}(\cdots)=E_{\overline g_r}[(\dots)^2]-[E_{\overline g_r}(\cdots)]^2$ for variances with respect to the distribution of the average propensities $\overline g_{r}$.

This decomposition of the variance of $k_r$ is convenient, as we know the distribution of $k_r$ for a given mean propensity $\overline{g}_r$. Specifically, $P(k_r|\overline{g}_r)$ is a Poisson distribution with mean $\overline{g}_r\Delta t$.
The variance of this Poisson distribution is also $\overline{g}_r\Delta t$. Therefore, the first term on the right-hand side of equation~(\ref{eq:tot_var}) evaluates to $E_{\overline g_r}(\overline g_r \Delta t)$. The second term simply relies on the mean of the Poisson distribution and reduces to $\mbox{var}_{\overline g_r}[\overline g_r \Delta t]$. Altogether we find
\be
\mbox{var}(k_r)= \Delta t E_{\overline g_r}({\overline g_r})+(\Delta t)^2 \mbox{var}_{\overline g_r}(\overline g_r).
\ee

\subsection{General chemical Langevin equation for any environment of discrete states}
Combining our results so far will turn our ansatz \eqref{eq:average_k_r} into
\begin{align}
n_i(t+\Delta t)&=n_i(t)+\sum_r \nu_{i,r} k_r(t) \nonumber \\
&=n_i(t) + \sum_r \nu_{i,r}\Delta t E_{\overline g_r}(\overline{g}_r) + \sum_r \nu_{i,r} \sqrt{E_{\overline g_r}(\overline{g}_r) + \Delta t \mbox{var}_{\overline g_r}(\overline g_r) } \sqrt{\Delta t}N_r(0,1), \label{eq:numerical_sde}
\end{align}
where $N_r(0,1)$ are Gaussian distributions with mean 0 and variance 1. The second term on the right hand side is due to the mean of the increments we calculated above, and the last term is contains our expression for the variance. 

We can interpret equation \eqref{eq:numerical_sde} as the Euler-Maruyama approximation of a stochastic differential equation. To do this, we note that  equation~\eqref{eq:gbardef} can be written as 
\begin{equation}
\overline{g}_r = \frac{1}{\Delta t}\int_t^{t+\Delta t} dt'\,g_{r,\sigma(t')}(t),
\end{equation}
i.e. $\overline g_r$ is the time-average of $g_{r,\sigma}(t)$ over the interval $[t,t+\Delta t$. Assuming that many environmental switches occur in $\Delta t$, and using the central limit theorem we then expect the variance of $\overline g_r$ to be proportional to $1/\Delta t$. Under these assumptions we define
\begin{equation}
q_r := \Delta t\mbox{var}_{\overline g_r}(\overline g_r).
\end{equation}
We then identify the stochastic differential equation
\begin{equation}
\frac{dn_i}{dt}
=\sum_r \nu_{i,r} E_{\overline g_r}(\overline{g}_r) + \sum_r  \sqrt{\nu_{i,r}^2E_{\overline g_r}(\overline{g}_r) + \nu_{i,r}^2 q_r} \xi_r,\label{eq:cle_0}
\end{equation}
where the $\xi_r$ denote independent Gaussian white noises.
\subsection{Chemical Langevin equation for an environment with two states, reflecting switching promoter states}
For an environment with two states (ON and OFF) and with switching rates $\lambda k_{\rm ON}$ from OFF to ON, and $\lambda k_{\rm OFF}$ from ON to OFF, the quantity $\mbox{var}_{\overline g_r}(\overline g_r)$ can be obtained using the method of \cite{hufton_pre_1} -- see in particular equation (25a) of that paper. In brief, one needs to find the variance of the time-averaged reaction propensity under random promoter switching, and in the limit that many switching events occur in $[t,t+\Delta t]$. We find
\be\label{eq:var_gbar}
\mbox{var}_{\overline g_r}(\overline g_r)\approx \frac{\theta^2}{\lambda \Delta t}(g_{r,{\rm ON}}-g_{r,{\rm OFF}})^2,
\ee
where $\theta^2=\frac{2k_{\rm ON}k_{\rm OFF}}{(k_{\rm ON}+k_{\rm OFF})^3}$. Hence, the quantity $q_r = \Delta t\mbox{var}_{\overline g_r}(\overline g_r)$ is indeed well-defined and
the chemical Langevin equation becomes
\BE
\frac{dn_i}{dt} &=& \sum_r \nu_{i,r} E_{\overline g_r}(\overline{g}_r) + \sum_r  \varsigma_{i,r}\xi_r(t),\label{eq:cle_1}
\EE
where the $\varsigma_{i,r}$ are defined as
\BE
\varsigma_{i,r}^2 = \nu_{i,r}^2E_{\overline g_r}(\overline{g}_r) + \frac{\nu_{i,r}^2 \theta^2}{\lambda}(g_{r,{\rm ON}}-g_{r,{\rm OFF}})^2, 
\EE

We note that the reaction propensities $g_{r,\sigma}$ typically scale with the system-size parameter $\Omega$ (the number of reactions of any one type in the system per unit time is proportional to the size of the system). Having this in mind, and writing $c_i(t)\equiv n_i(t)/\Omega$ for particle densities, and $g_{r,\sigma}=\Omega a_{r,\sigma}$ for the propensities we can re-write equation~(\ref{eq:cle_1}) in the form
\BE
\frac{dc_i}{dt} &=& \sum_r \nu_{r,i}E_{\overline g_r}(\overline{a}_r) +\sum_r  s_{i,r} \xi_r(t),\label{eq:cle_2}
\EE
where
\BE
s_{i,r}^2 = \frac{\nu_{i,r}^2}{\Omega}E_{\overline g_r}(\overline a_r(t)) + \frac{\nu_{i,r}^2}{\lambda}\theta^2 (a_{r,{\rm ON}}-a_{r,{\rm OFF}})^2. \label{eq:noise_contributions}
\EE
This makes it apparent that each reaction channel leads to two noise-contributions. The amplitude of one of the two contributions scales with $\Omega^{-1/2}$. This contribution represents copy-number noise. It has the same form as the noise contribution originally identified by Gillespie \cite{gillespieCLE}, except that Gillespie's reaction propensity is replaced by the average propensity over all promoter states.

The amplitude (standard deviation) of the second noise contribution in equation \eqref{eq:noise_contributions} scales with $\lambda^{-1/2}$. This contribution is not considered in Gillespie's derivation, and it represents noise due to switching between promoter states. 

\subsection{Chemical Langevin equation for the auto-negative feedback motif with promoter bursting}\label{sec:CLEsupp}


We now seek to express equation \eqref{eq:cle_2} for our model of an auto-negative feedback loop in the main paper (equation (1)). In this model, there are only two environmental states, ON and OFF. Switches from ON to OFF occur with reaction rate $\lambda [P(t)/(P_0)]^h$, and from OFF to ON with rate $\lambda$. Only the transcription reaction is affected by the environment. In the OFF state the rate for the reaction is zero, and in the ON state it is $\alpha_M$. The stationary probability to find the environment in the ON state is $1/[1+[P(t)/P_0]^h]$. Therefore the mean rate of transcription reactions that are {\em triggered} in the time interval between at time $t$ is
\be
E_{\overline{g}_r}(a_{\rm transcription})=\frac{\alpha_M}{1+[P(t)/(P_0(t))]^h}.
\ee
In our system, the product of a transcription rate only enters the system $\tau$ units of time later, which means that this rate of transcription does not affect the mRNA concentration at the current time, but at time $t+\tau$. The concentration of mRNA molecules entering the system at time $t$ was instead triggered at time $t-\tau$. Hence, the deterministic term reflecting transcription in equation \eqref{eq:cle_2} becomes
\be
E_{\overline{g}_r}(a_{\rm transcription})=\frac{\alpha_M}{1+[P(t-\tau)/(P_0(t))]^h}. \label{eq:mean_transcription}
\ee
It is a key advantage of the derivation of the chemical Langevin equation using increments of molecule numbers that this argument can be made. Much more elaborate arguments are necessary in the context of more formal Kramers--Moyal expansions \cite{brett}.

Next, we need to evaluate the noise term in equation \eqref{eq:cle_2}. The copy-number contribution of the transcription reaction is proportional to $E_{g_r}(a_{\rm transcription})$, provided in equation \eqref{eq:mean_transcription}. To derive the noise term representing promoter bursting, recall that $a_{\mathrm{transcription}, {\rm ON}}=\alpha_M$, and $a_{\mathrm{transcription}, \rm  OFF}=0$ for the transcription reaction. We also have $k_{\rm ON}=1$ and $k_{\rm OFF}= [P/(P_0)]^h$, defining these quantities as in \cite{hufton_pre_1}, which intentionally excludes $\lambda$. 
We arrive at
\be\label{eq:var}
\frac{\nu_{i,r}^2}{\lambda}\theta^2 (a_{r,{\rm ON}}-a_{r,{\rm OFF}})^2 = 
 \frac{2[P(t-\tau)/(P_0)]^h}{\left[1+(P(t-\tau/(P_0))^h\right]^3}\frac{1}{\lambda }\alpha_M^2.
\ee
Here, we have again considered that transcription noise at time $t$ originates at time $t-\tau$, due to the delay in the transcription reaction.

Putting everything together, we get for the densities $P=n_P/\Omega$ and $M=n_M/\Omega$ 
\BE
\frac{d}{dt} M &=& \frac{\alpha_M}{1+[P(t-\tau)/P_0]^h}-\mu_M M +\eta_M(t), \nonumber \\
\frac{d}{dt}P &=& \alpha_P M - \mu_P P +\eta_P(t) \label{eq:CLE_final}
\EE
with 
\BE
\avg{\eta_M(t)\eta_M(t')}&=&\Big\{\frac{\alpha_M}{\Omega}\frac{1}{1+[P(t-\tau)/P_0]^h}+\frac{\mu_M M}{\Omega}+\frac{\alpha_M^2}{\lambda}\frac{2[P(t-\tau)/P_0]^h}{\left(1+[P(t-\tau)/P_0]^h\right)^3}\Big\}\delta(t-t'), \nonumber \\
\avg{\eta_P(t)\eta_P(t')}&=&\left[\alpha_P M(t)+\mu_P P(t)\right]\delta(t-t'),\nonumber \\
\avg{\eta_M(t)\eta_P(t')}&=&0. \label{eq:noise_final}
\EE
Here, we have used a simplification introduced by Gillespie \cite{gillespieCLE}, which combines the noise terms for each chemical species into one Gaussian noise. Given that in our model each reaction contributes only to one chemical species (no reaction simultaneously changes the number of P and M molecules) the  random variables $\eta_M$ and $\eta_P$ independent.

\subsection{Spectra of fluctuations}\label{sec:spectra_sm}
Equation~(17) describes an Ornstein-Uhlenbeck process for $m$ and $p$. This process is linear in $m$ and $p$ by construction. To analyse this further, we move to Fourier space, and use $\omega$ to denote angular frequency. Fourier transforms are indicated by tildes, i.e. we have
\be\label{eq:FT}
\tilde y(\omega)=\frac{1}{\sqrt{2\pi}}\int dt\,e^{-i\omega t} y(t)
\ee
for the Fourier transform of $y(\cdot)$. The inverse transformation is
\be\label{eq:invFT}
 y(t)=\frac{1}{\sqrt{2\pi}}\int d\omega\,e^{i\omega t} \widetilde y(\omega).
\ee
We also have \be\label{eq:deltaexp}
\delta(\omega) = \frac{1}{2\pi}\int_{-\infty}^\infty dt ~ e^{i\omega t}.
\ee
From equation~(17) we then find
\BE
i\omega \widetilde m(\omega) &=& \alpha_M f'(P^*) e^{-i\omega\tau}\widetilde p(\omega)-\mu_M \widetilde m(\omega) +\widetilde \zeta_M(\omega), \nonumber \\
i\omega \widetilde p(\omega) &=& \alpha_P \widetilde m(\omega) - \mu_P \widetilde p(\omega) +\widetilde \zeta_P(\omega).\label{eq:lna_fourier}
\EE
Using equations~(\ref{eq:FT}), (19) and (\ref{eq:deltaexp}) we then find
\BE
\avg{\widetilde\zeta_M(\omega)\widetilde\zeta_M(\omega')}&=&\int \frac{dt\,dt'}{2\pi} e^{-i\omega t - i\omega' t'} \avg{\zeta_M(t)\zeta_M(t')} \nonumber \\
&=&\sigma_M^2 \int \frac{dt\,dt'}{2\pi} e^{-i\omega t - i\omega' t'} \delta(t-t') \nonumber \\
&=&\sigma_M^2 \int \frac{dt}{2\pi} \, e^{-i(\omega+\omega')t} = \sigma_M^2 \delta(\omega+\omega').
\EE
An analogous relation can be derived for $\avg{\widetilde\zeta_P(\omega)\widetilde\zeta_P(\omega')}$, and thus we have in summary,
\BE
\avg{\widetilde\zeta_M(\omega)\widetilde\zeta_M(\omega')}&=&\sigma_M^2\delta(\omega+\omega'), \nonumber \\
\avg{\widetilde\zeta_P(\omega)\widetilde\zeta_P(\omega')}&=&\sigma_P^2\delta(\omega+\omega'),\nonumber \\
\avg{\widetilde \zeta_M(\omega)\zeta_P(\omega')}&=&0.
\EE
Equations~(\ref{eq:lna_fourier}) can be written in matrix form as follows,
\be
\begin{pmatrix}
    i\omega+\mu_M & -\alpha_M f'(P^*)e^{-i\omega\tau} \\ -\alpha_P & i\omega+\mu_P
\end{pmatrix}
\begin{pmatrix}
    \widetilde m(\omega) \\\widetilde  p(\omega)
\end{pmatrix}
=
\begin{pmatrix}
    \widetilde \zeta_M(\omega) \\\widetilde  \zeta_P(\omega)
\end{pmatrix}.
\ee
Inverting the matrix on the left we then find
\BE
\begin{pmatrix}
    \widetilde m(\omega) \\\widetilde  p(\omega)
\end{pmatrix}
= \frac{1}{\Delta} \begin{pmatrix}
    i\omega+\mu_P & \alpha_M f'(P^*)e^{-i\omega\tau} \\ \alpha_P & i\omega+\mu_M
\end{pmatrix}
\begin{pmatrix}
    \widetilde \zeta_M(\omega) \\\widetilde  \zeta_P(\omega)
\end{pmatrix},
\EE
where
\BE
\Delta(\omega)&=&-\omega^2+i\omega(\mu_M+\mu_P)+\mu_M\mu_P -\alpha_M\alpha_P f'(P^*)e^{-i\omega\tau}\nonumber \\
&=&
\mu_M\mu_P-\alpha_M\alpha_pf'(P^*)\cos(\omega\tau)-\omega^2 + i\left[\omega(\mu_M+\mu_P)+\alpha_M\alpha_P f'(P^*)\sin(\omega\tau)\right].
\EE
We therefore arrive at
\BE\label{eq:mmpp}
\avg{\tilde m(\omega)\tilde m(\omega')}=S_M(\omega)\delta(\omega+\omega'), \nonumber \\
\avg{\tilde p(\omega)\tilde p(\omega')}=S_P(\omega)\delta(\omega+\omega'),
\EE
with
\BE
S_M(\omega)&=&\frac{(\omega^2+\mu_P^2)\sigma_M^2+[\alpha_M f'(P^*)]^2\sigma_P^2}{|\Delta(\omega)|^2},\nonumber \\
S_P(\omega)&=&\frac{\alpha_P^2\sigma_M^2+[\omega^2+\mu_M^2]\sigma_P^2}{|\Delta(\omega)|^2}.\label{eq:lna-powerSpectra}
\EE
The denominator in these expressions is
\BE
|\Delta(\omega)|^2&=&\left[\mu_M\mu_P-\alpha_M\alpha_Pf'(P^*)\cos(\omega\tau)-\omega^2\right]^2 +\left[\omega(\mu_M+\mu_P)+\alpha_M\alpha_P f'(P^*)\sin(\omega\tau)\right]^2.
\EE
\subsection{Variance of fluctuations}\label{sec:variance_derivation}
From equations~(\ref{eq:mmpp}) we have
\be
\avg{\widetilde p(\omega) \widetilde p(\omega')}=S_P(\omega)\delta(\omega+\omega').
\ee
Using this, and equation~(\ref{eq:invFT}), the variance $\avg{p(t)^2}$ is then obtained as
\BE
\avg{p(t)^2}&=&\frac{1}{2\pi} \int_{-\infty}^\infty d\omega \,d\omega'\, e^{i\omega t+i\omega' t} \avg{\widetilde p(\omega) \widetilde p(\omega')} \nonumber\\
&=& \frac{1}{2\pi} \int_{-\infty}^\infty d\omega S_P(\omega) = \frac{1}{\pi} \int_{0}^\infty d\omega S_P(\omega), 
\EE
where we have used the fact that $S_P$ is an even function of $\omega$.

\section{Modified model with alternative promoter kinetics}
\label{sec:promoter_alternative}
\subsection{Model definition}
\label{sec:modified_model_def}
In this section, we apply our theory to a modified version of the model in equation (1) of the main manuscript: Instead of allowing the protein to shorten the ON-period of the promoter, as proposed by \cite{alon2007}, we now parameterise the model such that the auto-repression modifies the OFF-duration of the promoter, as e.g. observed for the transcriptional repressor knirps in \textit{Drosophila} \cite{zhao}. We again design the promoter kinetics such that in the stationary limit we recover the Hill function in equation (10) of the main manuscript. This results in the reactions
\BE
G_{\text{ON}} &\stackrel{\lambda }{\longrightarrow}& G_{\text{OFF}}, \nonumber \\
G_{\text{OFF}} &\stackrel{\lambda (P/P_0)^{-h}}{\longrightarrow}& G_{\text{ON}}, \nonumber \\
M&\stackrel{\mu_M}{\longrightarrow}& \varnothing, \nonumber \\
P&\stackrel{\mu_P}{\longrightarrow}& \varnothing, \nonumber \\
\varnothing &\stackrel{\alpha_M \sigma}{\Longrightarrow}& M, \nonumber \\  
M&\stackrel{\alpha_P}{\longrightarrow}& M+P, \label{eq:full_modified_reactions}
\EE
where we use the same symbols as in equation (1) of the main paper. The difference compared to equation~(1) is in the promoter switching, where the protein concentration $P$ now  suppresses the OFF-to-ON reaction, as opposed to enhancing the ON-to-OFF reaction. This means that $\lambda (P/P_o)^h$ in the original setup (1) has been replaced by $\lambda$, and what was $\lambda$ in the old setup of equation (1) of the main manuscript has been replaced by $\lambda (P/P_0)^{-h}$. Hence, effectively, this is a re-scaling of $\lambda$ in equation (1) of the main manuscript by a factor of $(P/P_0)^{-h}$.

\subsection{Limiting case for infinite system size and the fully deterministic limit}
Given that the reactions involving mRNA and protein are unchanged, the limit $\Omega \rightarrow \infty$, in which copy-number noise is neglected, leads to an expression that is identical to equation (13) of the main manuscript, which we obtained for the original model in equation (1), also in the main manuscript. A difference arises from the state-variable $\sigma(t)$, which is now governed by the new promoter dynamics in equation \eqref{eq:full_modified_reactions} instead of those in (1). Specifically $\sigma(t)$ will undergo the transition $1\rightarrow0$ with rate $\lambda$ and the transition $0\rightarrow 1$ with rate $\lambda (P/P_0)^{-h}$.

The fully deterministic limit, which is mathematically emerging from $\lambda\rightarrow \infty$ jointly with $\Omega\rightarrow\infty$, and in which both bursting noise and copy-number noise are neglected, is identical to equation (12). Specifically, the probability of the promoter being in the ON state is now
\BE\label{eq:p_unbound2}
p_{\rm ON}=\frac{\lambda (P/P_0)^{-h}}{\lambda [1+(P/P_0)^{-h}]}=\frac{1}{1+(P/P_0)^h}=:f(P),
\EE
which is the same as in equation (9). The deterministic rate equations (12) are therefore unchanged by moving from the model in (1) to the alternative model in equation \eqref{eq:full_modified_reactions}.

\subsection{Chemical Langevin equation}\label{sec:CLEsupp2}
Following on from Eq.~(\ref{eq:p_unbound2}) the mean transcription rate $E_{\overline{g}_r}(a_{\rm transcription})$ remains as in the original setup, and is given by equation (\ref{eq:mean_transcription}). Regarding the noise term in equation \eqref{eq:cle_2}, the copy-number contribution of the transcription reaction is proportional to $E_{g_r}(a_{\rm transcription})$ in \eqref{eq:mean_transcription}, and is thus unchanged compared to the original setup.

To derive the noise term representing bursting effects, we use that, as before,  $a_{\mathrm{transcription}, {\rm ON}}=\alpha_M$, and $a_{\mathrm{transcription}, \rm  OFF}=0$. We now have $k_{\rm ON}=(P/P_0)^{-h}$ and $k_{\rm OFF}= 1$. The next steps are again to calculate the quantity 
\be
\theta^2=\frac{2k_{\rm ON}k_{\rm OFF}}{(k_{\rm ON}+k_{\rm OFF})^3},
\ee
and then to use Eq.~(\ref{eq:var_gbar}).
We find
\BE
\theta^2&=&\frac{2k_{\rm ON}k_{\rm OFF}}{(k_{\rm ON}+k_{\rm OFF})^3} \nonumber \\
&=&\frac{1}{\lambda}\frac{(P/P_0)^{2h}}{[1+(P/P_0)^{h}]^3}
\EE
This results in
\be\label{eq:var2}
\frac{\nu_{i,r}^2}{\lambda}\theta^2 (a_{r,{\rm ON}}-a_{r,{\rm OFF}})^2 = 
 \frac{2[P(t-\tau)/(P_0)]^{2h}}{\left[1+(P(t-\tau/(P_0))^{h}\right]^3}\frac{1}{\lambda }\alpha_M^2
\ee
for the noise term representing bursting effects.
We highlight that this term is notably different from equation~(\ref{eq:var}).

\medskip

The CLE for the concentrations $P=n_P/\Omega$ and $M=n_M/\Omega$ for our alternative model in equation \eqref{eq:full_modified_reactions} is then
\BE
\frac{d}{dt} M &=& \frac{\alpha_M}{1+[P(t-\tau)/P_0]^h}-\mu_M M +\eta_M(t), \nonumber \\
\frac{d}{dt}P &=& \alpha_P M - \mu_P P +\eta_P(t), \label{eq:CLE_final2}
\EE
with 
\BE
\avg{\eta_M(t)\eta_M(t')}&=&\Big\{\frac{\alpha_M}{\Omega}\frac{1}{1+[P(t-\tau)/P_0]^h}+\frac{\mu_M M}{\Omega}+\frac{\alpha_M^2}{\lambda}\frac{2[P(t-\tau)/P_0]^{2h}}{\left(1+[P(t-\tau)/P_0]^{h}\right)^3}\Big\}\delta(t-t'), \nonumber \\
\avg{\eta_P(t)\eta_P(t')}&=&\frac{1}{\Omega}\left[\alpha_P M(t)+\mu_P P(t)\right]\delta(t-t'),\nonumber \\
\avg{\eta_M(t)\eta_P(t')}&=&0. \label{eq:noise_final2}
\EE
The key difference compared to the expressions in Eq.~(\ref{eq:noise_final}) is an additional factor of  $[P(t-\tau)/P_0]^{h}$ in the second term in the curly brackets in the expression for $\avg{\eta_M(t)\eta_M(t')}$, changing the numerator $2[P(t-\tau)/P_0]^{h}$ to $2[P(t-\tau)/P_0]^{2h}$.

\subsection{Linear-noise approximation}
For the linear-noise approximation we obtain an identical expression to equation 
(17), where the noise terms follow the expressions in equation (19), except that their magnitude is now given by 
\BE
\sigma_M^2&=&\frac{1}{\Omega}\left[\alpha_M f(P^*)+\mu_M M^*\right] + \frac{1}{\lambda}\frac{2[P^*/P_0]^{2h}}{\left(1+[P^*/P_0]^{h}\right)^3}\alpha_M^2,\nonumber \\
\sigma_P^2&=&\frac{1}{\Omega}\left[\alpha_P M^*+\mu_P P^*\right],\label{eq:sigma2}
\EE
where, compared to equation~(20), the second term in $\sigma_M^2$ has an additional factor of $[P^*/P_0]^h$.
Conveniently, the fixed-point concentrations $M^*$ and $P^*$ remain unchanged compared to the original setup.

The calculation of the spectra of fluctuations about the fixed point follows the same steps as in Sec.~\ref{sec:spectra_sm}. 
The resulting formula is identical to equations \eqref{eq:mmpp} and (21), except that the expressions in \eqref{eq:sigma2} need to be used for $\sigma_M^2$ and $\sigma_P^2$.  

\subsection{A scaling relation exists between the bursting parameters of both models}
\label{sec:supp_scaling}
We note that $\sigma_M^2$ is the only place where the bursting parameter $\lambda$ appears in the LNA, and in the resulting spectra of fluctuations. Importantly, the new noise terms of the LNA in equation \eqref{eq:sigma2} can be obtained from the previous expressions of the original model in equation (20) of the main manuscript by substituting
\be
\lambda \to \lambda \left(\frac{P^*}{P_0}\right)^{-h}.
\ee
This also follows from the definition in equation \eqref{eq:full_modified_reactions}, and using the LNA assumption that $P(t)$ can be replaced by $P^*$ in all reaction rates. Hence, we expect the effects of bursting noise between the two models to have the same magnitude if $\lambda$ is rescaled by $(P_0/P^*)^h$.

\subsection{Stochastic differential equations approximate the alternative model well}
\begin{figure}[t!]
    \centering
    \includegraphics{ 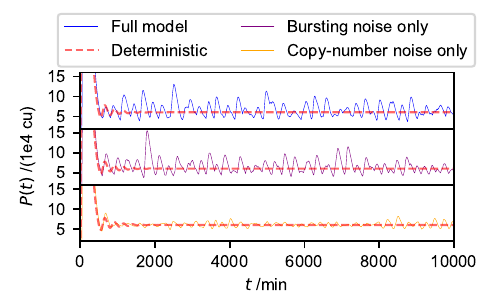
    }
    \caption{\textbf{Noise-induced oscillations under the alternative model design}. 
    'Full model' refers to  equation~\eqref{eq:full_modified_reactions}, `bursting noise only' refers to equation (13) simulated using the transition rates for $\sigma$ from equation \eqref{eq:full_modified_reactions}, i.e. the approximation valid in the limit $\Omega\to\infty$ and $\lambda$ remaining finite. `Copy-number noise only' refers to equation (11) for infinitely fast transcriptional burst frequencies ($\lambda\to\infty$, $\Omega$ finite). The deterministic model refers to equation (12), valid in the combined limit $\Omega\to\infty$ and $\lambda\to\infty$. Parameters are as in figure 2. We use $\Omega = 1$ for the full model and the copy-number noise only model, whereas the remaining models assume $\Omega=\infty$. Similarly, we use $\lambda = 5$min$^{-1}$ for the full model and the bursting-noise only model, where the remaining models assume $\lambda=\infty$. Details of the computational implementation are available in supplementary section \ref{sec:supp_numerics}.}\label{fig:alternative_noise}
\end{figure}

The  modified model in equation \eqref{eq:full_modified_reactions} behaves qualitatively similar to our original model in equation (1) of the main manuscript. Importantly, the model also exhibits oscillations that are induced by noise, and amplified by bursting noise (figure \ref{fig:alternative_noise}, cf. figure 2). In this figure, we chose all parameters identical to those in figure 2, except for the bursting parameter, for which we chose $\lambda = 5$min$^{-1}$ instead of 0.01min$^{-1}$. This value was chosen such that oscillation amplitudes are comparable to those in figure 2.

Changing the bursting parameter $\lambda$ is necessary due to the scaling relationship mentioned in the supplementary section \ref{sec:supp_scaling}. We provide a numerical evaluation of that scaling by considering the OFF-periods of the promoter in both models. The OFF-periods between the two states of the promoter in the modified model are set by the reaction rate $\lambda(P/P_0)^h$, whereas it is $\lambda$ in the original model. Hence, OFF-periods are of the duration of $(P_0/P)^h/\lambda$ instead of $1/\lambda$. In our simulations, the steady state of the deterministic equations is at around $P=60000$cu (figures \ref{fig:alternative_noise} and 2), with $P_0$ chosen at 24201.01cu, and with $h=4.78$. This gives a scaling factor of around 0.013 between off-durations of the alternative model \eqref{eq:full_modified_reactions} and those of the original model (1), illustrating that choosing different values of $\lambda$ is essential when comparing bursting effects between the models.

\begin{figure}
\includegraphics{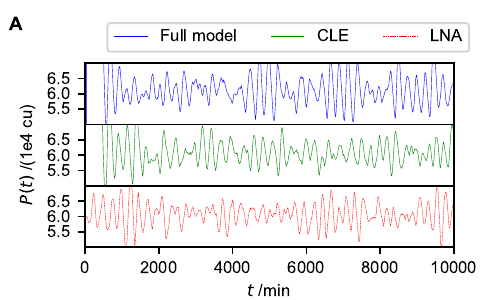}
\includegraphics{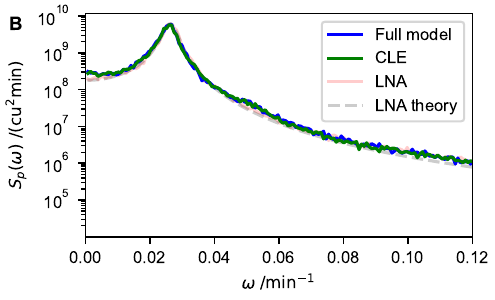}
    \caption{\textbf{Stochastic differential equations approximate the alternative model well}. 
    For the following models we show individual trajectories in panel A and power spectra in panel B.
    The full model is a Gillespie simulation of equation~\eqref{eq:full_modified_reactions}, `CLE' stands for chemical Langevin equation (see equation~\eqref{eq:CLE_final2}), and LNA stands for `linear noise approximation' (equation~(17)), simulated using the definition of $\sigma_M^2$ from equation \eqref{eq:sigma2} instead of equation (20). `LNA theory' refers to the analytically obtained power spectrum in equation~(21), also calculated using $\sigma_M^2$ from equation \eqref{eq:sigma2}. The parameters used are the same as in figure~2, except that here $\Omega=100$ and $\lambda=100\mathrm{min}^{-1}$. All three models exhibit noise-induced oscillations and their power spectra match. Note that $\omega$ represents angular frequencies. To obtain the power spectra, 100 trajectories of each model are generated. They are simulated for $T=10000$min and the first 200 minutes are discarded to account for equilibration. Further details of the computational implementation are available in supplementary section \ref{sec:supp_numerics}.\label{fig:alternative_cle_test}}
\end{figure}

Having established that the alternative model behaves similarly to the original model in inducing oscillations, we investigate to what extent our extended chemical Langevin equations are able to approximate the effects of bursting noise in the new model. We find that, indeed, the new chemical Langevin equations \eqref{eq:CLE_final2} predict timeseries that are qualitatively similar to those of the full model, in the sense that they predict oscillations of comparable amplitude and frequency (figure \ref{fig:alternative_cle_test}A). The same is true for the linear-noise approximation of the new model (also figure \ref{fig:alternative_cle_test}A). When comparing the power spectra of the three models (the full model, the chemical Langevin equation, and the linear-noise approximation, figure \ref{fig:alternative_cle_test}B) we find that, similar to figure 3B of the main manuscript, the power spectra of all three models agree closely with each other.

\section{Computational implementation}\label{sec:supp_numerics}
In this section, we describe how we simulate the models in equations (1), (12), (13), (14), and (17).\\

\noindent
\textbf{Full model in equation (1)}\\
The chemical reaction system in equation (1) of the main manuscript is simulated using the Gillespie-rejection algorithm reported by \cite{cai}. Simulated numbers of mRNA and protein molecules are recorded every minute for plotting and further analysis.\\

\noindent
\textbf{Deterministic model in equation (12)}\\
The deterministic model (12) is solved using a Forward-Euler approximation and a timestep of $\Delta t = 0.01$.\\

\noindent
\textbf{Promoter-noise only model in equation (13)}\\
The piecewise-deterministic Markov process, i.e. the model emerging in the limit of infinite copy numbers (13), is simulated using an adapted Forward-Euler scheme. Specifically, at each timestep starting at time $t$, the concentration of mRNA and protein are propagated to $t+\Delta t$ using the Forward-Euler method according to the promoter state $\sigma$ at time $t-\tau$, i.e. $\Delta t$ is the time increment used in the Forward-Euler approximation. After propagating the mRNA and Protein concentrations $M$ and $P$ in this way, the propensity to switch from the current promoter state is calculated. This propensity is $g=\lambda$ if the promoter is currently in the OFF state. Otherwise, it is $g = \lambda (P(t)/P_0)^n$. Then, a random number $\Delta t_r$ is drawn from an exponential distribution with mean $1/g$. If the generated time $\Delta t_r$ is less than the timestep $\Delta t$, then the promoter state at $t+\Delta t$ is changed from its current state into the second available promoter state (i.e. from OFF to ON or from ON to OFF). This simulates that the promoter switches to the opposite state with probability $1-e^{-g\Delta t}$, and remains in the current state with probability $e^{-g\Delta t}$. This procedure is repeated for the duration $T$ of the simulation. A timestep of $\Delta t=0.0001$ is used in our simulations, and values of $M$ and $P$ at every full minute are recorded for plotting.\\

\noindent
\textbf{Extended chemical Langevin equation in equation (14)}\\
The chemical Langevin equation (14) is simulated using a delayed Euler--Maruyama scheme. We use reflecting boundary conditions on the copy numbers of the protein and the mRNA at zero copy numbers, i.e. if the predicted number of protein or mRNA is negative at a given timestep, the absolute value of that prediction is used when propagating into the next timestep. A timestep of $\Delta t=0.01$ is used in the approximation, and values of $M$ and $P$ at every full minute are recorded for plotting.\\

\noindent
\textbf{Linear-noise approximation in equation (17)}\\
The linear-noise approximation (17) is simulated in the same way as the chemical Langevin equation, except that we do not use a reflecting boundary condition at zero. This difference is necessary to enable comparison between analytical and computational predictions from the linear-noise approximation.\\

\noindent
\textbf{Condition for past times}\\
At past times ($t<0$), all models except the linear-noise approximation assume $M(t)=0$ and $P(t)=0$. The linear-noise approximation assumes instead $M(t)=M^*$ and $P(t)=P^*$, where $M^*$ and $P^*$ are the concentrations of mRNA and protein at the steady state, respectively. This is equivalent to saying that $m(t) = p(t) = 0$. Defining values for past times in this way is
required to interrogate $P(t-\tau)$ when $t<\tau$.\\ 

\noindent
\textbf{Calculation of power spectra}\\
To calculate power spectra, $N_{\mathrm{pow}}$ trajectories of a model are calculated for the duration $T_{\mathrm{pow}}$ after an initial equilibriation period of $T_\mathrm{eq}$. For each trajectory, the Python numpy fast Fourier transform routine is used to calculate the discrete Fourier transform $\tilde{P}_\mathrm{disc}(\omega)$ of the protein concentration. Then, the average of the square magnitude of this Fourier transform $|\tilde{P}_\mathrm{disc}(\omega)|^2$ is taken over all simulated trajectories. As this value represents the power under the discrete Fourier transform, we convert it into the Power spectral density used in this paper using the conversion
\begin{equation}
S_P(\omega) = E(|\tilde{P}_\mathrm{disc}(\omega)|^2)2T_\mathrm{pow}/n_t^2
\end{equation}
where $n_t$ is the number of timesteps comprising the duration $T_\mathrm{pow}$. Since protein concentrations in all models are only recorded every minute, $n_t$ is simply the duration of $T_\mathrm{pow}$ in minutes.\\

\noindent
\textbf{Code availability}\\
Each algorithm is implemented in python and the code to simulate each model is available under the Zenodo repository \url{https://zenodo.org/records/15046770}.

\end{document}